\documentclass[prx, aps, showpacs, twocolumn, preprintnumbers, amsmath, amssymb, superscriptaddress, nofootinbib, longbibliography, nofootinbibjustification=justified, singlelinecheck=false, 10pt]{revtex4-2}

\usepackage[colorlinks=true, linkcolor=blue, citecolor=blue, urlcolor=blue]{hyperref}
\usepackage{amsthm}
\theoremstyle{plain}
\newtheorem*{conjecture}{Conjecture}
\newtheorem*{question}{Question}
\newtheorem*{theorem}{Theorem}
\newtheorem*{lemma}{Lemma}
\newtheorem*{remark}{Remark}
\usepackage{xcolor}
\usepackage{graphicx}
\usepackage{tikz}
\usetikzlibrary{arrows.meta,positioning,calc,fit,backgrounds}
\usepackage{booktabs}
\newcommand{\ket}[1]{|#1\rangle}

\usepackage{listings}

\definecolor{leanKw}{HTML}{225BB4}
\definecolor{leanType}{HTML}{8C45A5}
\definecolor{leanIdent}{HTML}{117B78}
\definecolor{leanComment}{HTML}{667085}
\definecolor{leanBg}{HTML}{F7F9FC}
\definecolor{leanRule}{HTML}{D9E1EA}

\lstdefinelanguage{LeanFour}{
  sensitive=true,
  alsoletter={_},
  morekeywords={abbrev,def,theorem,fun,if,then,else},
  morekeywords=[2]{Nat,Type,Prop,Semiring,Fin,List,Even},
  morekeywords=[3]{
    WeightsN,EdgeN,pmSumListAux,pmSumList,
    pmSumN,EqSystemN,mkEdge,allEqual,vertices,
    eqSystem_no_solution_ge6_ge3,
    eqSystem_no_solution_even_ge4_d_eq_n_explicit
  },
  keywordstyle=\color{leanKw}\bfseries,
  keywordstyle=[2]\color{leanType},
  keywordstyle=[3]\color{leanIdent},
  morecomment=[s]{/-}{-/},
  morecomment=[l]{--},
  commentstyle=\color{leanComment}\itshape,
  literate=
    {α}{{\ensuremath{\alpha}}}1
    {ι}{{\ensuremath{\iota}}}1
    {∀}{{\ensuremath{\forall}}}1
    {∃}{{\ensuremath{\exists}}}1
    {¬}{{\ensuremath{\neg}}}1
    {→}{{\ensuremath{\to}}}1
    {≥}{{\ensuremath{\geq}}}1
    {ℂ}{{\ensuremath{\mathbb{C}}}}1
}

\lstdefinestyle{leanpaper}{
  language=LeanFour,
  basicstyle=\ttfamily\fontsize{7.1}{9.2}\selectfont,
  backgroundcolor=\color{leanBg},
  frame=single,
  framerule=0.3pt,
  rulecolor=\color{leanRule},
  framesep=4pt,
  columns=fullflexible,
  keepspaces=true,
  showstringspaces=false,
  breaklines=true,
  breakatwhitespace=false,
  aboveskip=7pt,
  belowskip=7pt,
  xleftmargin=1pt,
  xrightmargin=1pt
}

\begin{document}

\title{A Tensor-Algebraic No-Go Theorem\\for High-Dimensional Photonic Greenberger--Horne--Zeilinger States \\Discovered with AlphaProof Nexus}

\author{Mario Krenn}
\affiliation{Department of Computer Science, Faculty of Science, University of Tübingen, Tübingen, Germany}

\author{Moritz Firsching}
\affiliation{Google DeepMind, Zürich, CH}

\author{George Tsoukalas}
\affiliation{Google DeepMind, London, UK}

\author{Rishikesh Gajjala}
\affiliation{Center for Quantum and Topological Systems, NYU Abu Dhabi, Abu Dhabi, UAE}

\author{Xuemei Gu}
\affiliation{Institut f\"{u}r Festk\"{o}rpertheorie und Optik, Friedrich-Schiller-Universit\"{a}t Jena, Jena, Germany}

\author{Swarat Chaudhuri}
\affiliation{Google DeepMind, London, UK}

\begin{abstract}
Photonic GHZ states generated with probabilistic pair sources can be described by edge-coloured, complex-weighted graphs.  In this representation, conditioning on an event with one photon in every detector turns the output amplitudes into coherent sums over perfect matchings, and generating an \(N\)-photon \(D\)-dimensional GHZ state is equivalent to constructing a monochromatic quantum graph. Here, we prove that no complex-weighted monochromatic quantum graph exists for any even \(N\ge4\) when \(D\ge N\).  The proof, discovered through AI-assisted formal proof search in Lean using AlphaProof Nexus, constructs a weighted sum \(\mathcal F\) of amplitudes whose coefficients factor vertex by vertex.  Summing over the colours at one fixed vertex reduces the problem to an \((N-1)\times N\) matrix of effective edge weights.  For the graph amplitudes \(\omega\) and GHZ target \(\Delta\), the row--column imbalance yields \(\mathcal F(\omega)=0\) but \(\mathcal F(\Delta)\neq0\), contradicting the required equality \(\omega=\Delta\). Finally, we give a physical interpretation of the AI-discovered mathematical proof.
\end{abstract}

\maketitle

\section*{Introduction}
High-dimensional multipartite entanglement is an increasingly important resource in quantum information science~\cite{erhard2020advances, malik2026high}. Compared with the qubit case, increasing the local dimension
enlarges the joint state space rapidly with both the number of particles and the local dimension, yielding a richer structure of quantum correlations and new capabilities across a range of applications. These include
high-dimensional quantum communication~\cite{cozzolino2019high,yu2025quantum}, high-dimensional quantum teleportation~\cite{luo2019quantum,hu2020teleportation}, and quantum computation~\cite{reimer2019high,lib2024resource, ringbauer2022universal, chi2022programmable}. High-dimensional entanglement also raises fundamental questions about quantum correlations that have no direct analogue in bipartite or qubit settings, including the existence and classification of absolutely maximally entangled states~\cite{huber2017absolutely,rather2022thirty} and all-versus-nothing violations of local realism~\cite{lawrence2014rotational,lawrence2017mermin}.

Photons are natural carriers of high-dimensional quantum information, since their orbital angular momentum, path, time-bin, frequency, and other internal degrees of freedom provide multi-level encodings within a single particle. Entangling such encodings across photons gives access to high-dimensional multipartite states. Multi-photon high-dimensional entanglement was first demonstrated using orbital angular momentum~\cite{malik2016multi} and path~\cite{hu2020layered}, followed by high-dimensional Greenberger--Horne--Zeilinger states realised with orbital angular momentum~\cite{erhard2018experimental} and path encoding~\cite{bao2023very,hu2025observation}. These experiments generated three-dimensional ($D=3$) GHZ states of three and four particles (\(N=3\) and \(N=4\)), respectively.
These experiments demonstrate that small-scale high-dimensional multipartite entanglement can be generated in photonic systems and motivate a general question:

\textit{Which high-dimensional multipartite entangled states can be generated systematically with standard photonic resources?}

This question can be formulated precisely using a link between quantum optics and graph theory~\cite{krenn2017quantum,gu2019graph2,gu2019graph3}: it asks whether a system of \(D^N\) polynomial equations over \(\mathbb C\) in \(D^2N(N-1)/2\) complex variables has a solution. The system has solutions for \(D=2\) and arbitrary even \(N\), because photonic qubit GHZ states can be constructed~\cite{lu2007experimental, yao2012observation,wang2016experimental,zhong201812}; for four photons, qutrit GHZ states can also be created~\cite{erhard2018experimental,bao2023very,hu2025observation}. Despite extensive numerical searches, no solutions beyond these constructions are known. Theoretical work has shown that no other GHZ states can be generated under restricted-resource assumptions, such as the absence of multipartite interference~\cite{bogdanov2017mathoverflow} or restricted correlations~\cite{chandran2024krenngu}. However, no proof has established the existence or nonexistence of higher-dimensional GHZ states when the full capabilities of linear quantum optics~\cite{pan2012multiphoton} and probabilistic pair generation from nonlinear sources~\cite{hochrainer2022quantum} are available.

Here, we give the first unconditional proof that no monochromatic quantum graph exists when \(D\geq N\), for every even \(N\geq4\). The proof was discovered by \textit{AlphaProof Nexus} \cite{tsoukalas2026advancing}, an agentic AI system which autonomously searches for formalised proofs in Lean \cite{demoura2021lean}.

We first review the connection between photonic experiments and graph theory, then present the proof, and finally discuss its physical interpretation.

\section*{Background}
\subsection*{Photonic GHZ states and graphs}

Photonic Greenberger--Horne--Zeilinger (GHZ) states are multipartite entangled states of the form
\begin{equation}
\ket{\mathrm{GHZ}_{N,D}}
=
\frac{1}{\sqrt D}
\sum_{j=0}^{D-1}
\ket{j}^{\otimes N},
\label{eq:ghz-state}
\end{equation}
where \(N\) is the number of photons and \(D\) is the local dimension of each photon. We write \(C_D:=\{0,\ldots,D-1\}\) for the full colour set. To construct a proof, we use the graph-theoretic description of linear-optical quantum experiments with probabilistic photon-pair sources, which are among the most common experimental building blocks in photonic quantum optics laboratories. Such experiments can be represented by edge-coloured and edge-weighted graphs~\cite{krenn2017quantum,gu2019graph2,gu2019graph3}.

In this representation, the vertices of the graph correspond to photonic paths, or equivalently to the locations at which photons may be detected. An edge between two vertices represents the creation of a correlated photon pair in the corresponding two paths. The colours assigned to the edge encode the mode numbers of the photons. For example, a photon pair in the state \(\ket{0,0}\) may be represented by a blue edge, a pair in the state \(\ket{1,1}\) by a red edge, and a pair in the state \(\ket{2,2}\) by a green edge. Photon pairs can also have different mode numbers, such as \(\ket{0,2}\). In this case, the corresponding edge is bi-coloured: the endpoint incident to the first vertex has the colour associated with mode \(0\), while the endpoint incident to the second vertex has the colour associated with mode \(2\).

In addition to colours, every edge \(e\) carries a complex-valued weight \(w_e\), which represents the magnitude and phase of the corresponding two-photon amplitude. Thus, the graph encodes both which photon pairs can be produced and with which complex amplitudes they are produced. Conversely, every such edge-coloured and edge-weighted graph can be translated into a linear-optical experiment with probabilistic photon-pair sources.

We now condition on the event that exactly one photon is detected in each photonic path. In the graph picture, this corresponds to selecting perfect matchings of the graph. A perfect matching is a subgraph in which every vertex is incident to exactly one edge. Hence, each perfect matching describes one possible way of producing exactly one photon at every vertex.

Each perfect matching \(M\) contributes an amplitude to a computational-basis state. Since every vertex is incident to exactly one edge in a perfect matching, each vertex inherits a unique colour $c_{v_i}$ from the endpoint of that edge. This assignment of colours to vertices is called the \emph{inherited vertex colouring} of the perfect matching \(M\) and is denoted by $\operatorname{IVC}(M)$; explicitly,
\begin{equation}
\operatorname{IVC}(M)
=
(c_{v_1},\ldots,c_{v_N}),
\label{eq:inherited-vertex-coloring}
\end{equation}
The perfect matching \(M\) contributes to the computational-basis state \(\ket{c_{v_1},\ldots,c_{v_N}}\) with amplitude equal to the product of the weights of all edges in \(M\):
\begin{equation}
w_M=\prod_{e\in M} w_e.
\label{eq:perfect-matching-weight}
\end{equation}

Importantly, different perfect matchings may lead to the same inherited vertex colouring. Physically, that means that several distinct physical processes may contribute to the same quantum-state term. The final amplitude of a given term is therefore obtained by coherently summing the contributions of all perfect matchings that induce the same inherited vertex colouring. This coherent sum permits constructive and destructive interference between different perfect matchings.

Formally, let \(G\) be an edge-coloured, complex-weighted multigraph with \(N\) vertices, with each edge endpoint coloured by an element of \(C_D\), and with each edge \(e\) carrying a weight \(w_e\in\mathbb C\). Let \(\mathcal{PM}(G)\) denote the set of edge-level perfect matchings of \(G\). Write
\begin{equation}
\begin{aligned}
V(G)&=\{v_1,\ldots,v_N\},\\
\mathbf c&=(c_{v_1},\ldots,c_{v_N})
\in C_D^N.
\end{aligned}
\label{eq:coloring-domain}
\end{equation}
These are, respectively, the vertex set and a possible inherited vertex colouring. The perfect matchings inducing \(\mathbf c\) form the set
\begin{equation}
\mathcal{PM}_{\mathbf c}(G)
=
\left\{
M\in \mathcal{PM}(G)
:
\operatorname{IVC}(M)=\mathbf c
\right\},
\label{eq:matchings-with-coloring}
\end{equation}
An element of \(\mathcal{PM}(G)\) selects specific edges, including a specific choice among parallel edges. The amplitude, or weight, of the quantum-state term with colouring \(\mathbf c\) is the sum of the weights of all matchings inducing \(\mathbf c\):
\begin{equation}
\omega_{\mathbf c}
=
\omega_{c_{v_1},\ldots,c_{v_N}}
=
\sum_{M\in \mathcal{PM}_{\mathbf c}(G)}
w_M.
\label{eq:coloring-amplitude}
\end{equation}

\begin{figure*}[t]
\centering
\resizebox{\textwidth}{!}{%
\begingroup

\definecolor{ghzblue}{HTML}{1C78C0}
\definecolor{ghzred}{HTML}{C93640}
\definecolor{ghzgreen}{HTML}{259A63}
\definecolor{ghzorange}{HTML}{EC921C}
\definecolor{ghzink}{HTML}{24272B}
\definecolor{ghzmuted}{HTML}{656B70}
\definecolor{ghzrule}{HTML}{D7DBDF}
\tikzset{
  ghz edge/.style={line width=0.62pt,line cap=butt,line join=round},
  ghz example edge/.style={line width=1.35pt,line cap=round},
  ghz vertex/.style={circle,draw=ghzink,fill=white,line width=0.85pt,
                    minimum size=4.8mm,inner sep=0pt,font=\fontsize{11}{12}\selectfont},
  ghz small vertex/.style={circle,draw=ghzink,fill=white,line width=0.65pt,
                    minimum size=3.4mm,inner sep=0pt,font=\fontsize{8}{9}\selectfont},
  ghz header/.style={anchor=west,font=\sffamily\bfseries\fontsize{9.5}{11.2}\selectfont,
                    text=ghzink,inner sep=0pt},
  ghz text/.style={font=\fontsize{9}{11}\selectfont,text=ghzink,inner sep=0pt},
  ghz small text/.style={font=\fontsize{8}{10}\selectfont,text=ghzmuted,inner sep=0pt}
}
\newcommand{\GHZhalfedge}[9]{%
  \draw[ghz edge,#8] (#1) .. controls (#2) and (#3) .. (#4);
  \draw[ghz edge,#9] (#4) .. controls (#5) and (#6) .. (#7);
}

\begin{tikzpicture}[x=1cm,y=1cm]
\path[use as bounding box] (-0.10,0.23) rectangle (17.80,-8.55);
\draw[ghzrule,line width=0.45pt] (8.05,0.05) -- (8.05,-8.13);
\draw[ghzrule,line width=0.45pt] (0,-3.13) -- (7.62,-3.13);

\node[ghz header] at (0,0) {(a)\quad Graph--experiment correspondence};
\node[ghz small vertex] at (1.00,-0.70) {$r$};
\node[ghz text] at (2.40,-0.70) {$\longleftrightarrow$};
\node[ghz text,anchor=west] at (2.97,-0.70) {Photonic path / detector};

\draw[ghz example edge,ghzblue] (0.20,-1.45) -- (1.00,-1.45);
\draw[ghz example edge,ghzorange] (1.00,-1.45) -- (1.80,-1.45);
\node[ghz small vertex] at (0.20,-1.45) {$r$};
\node[ghz small vertex] at (1.80,-1.45) {$s$};
\node[ghz text] at (1.00,-1.15) {$w_e$};
\node[ghz text] at (2.40,-1.45) {$\longleftrightarrow$};
\node[ghz text,anchor=west] at (2.97,-1.45) {Photon-pair term $w_e\ket{0}_r\ket{3}_s$};

\node[ghz text] at (1.00,-2.13) {$w_e\in\mathbb C$};
\node[ghz text] at (2.40,-2.13) {$\longleftrightarrow$};
\node[ghz text,anchor=west] at (2.97,-2.13) {Amplitude and phase};

\node[ghz small text,anchor=west] at (0,-2.73) {Internal mode:};
\draw[line width=1.75pt,ghzblue] (2.110000,-2.730000) -- (2.550000,-2.730000);
\node[ghz text,anchor=west] at (2.650000,-2.730000) {$0$};
\draw[line width=1.75pt,ghzred] (3.360000,-2.730000) -- (3.800000,-2.730000);
\node[ghz text,anchor=west] at (3.900000,-2.730000) {$1$};
\draw[line width=1.75pt,ghzgreen] (4.610000,-2.730000) -- (5.050000,-2.730000);
\node[ghz text,anchor=west] at (5.150000,-2.730000) {$2$};
\draw[line width=1.75pt,ghzorange] (5.860000,-2.730000) -- (6.300000,-2.730000);
\node[ghz text,anchor=west] at (6.400000,-2.730000) {$3$};

\node[ghz header] at (0,-3.58) {(b)\quad Coincidences and perfect matchings};
\draw[ghz example edge,ghzblue] (0.430000,-4.330000) -- (1.670000,-4.330000);
\draw[ghz example edge,ghzblue] (0.430000,-5.570000) -- (1.670000,-5.570000);
\draw[ghz example edge,ghzred] (0.430000,-4.330000) -- (0.430000,-5.570000);
\draw[ghz example edge,ghzred] (1.670000,-4.330000) -- (1.670000,-5.570000);
\node[ghz small vertex] at (0.430000,-4.330000) {$1$};
\node[ghz small vertex] at (1.670000,-4.330000) {$2$};
\node[ghz small vertex] at (0.430000,-5.570000) {$3$};
\node[ghz small vertex] at (1.670000,-5.570000) {$4$};
\draw[ghz example edge,ghzblue] (3.270000,-4.330000) -- (4.510000,-4.330000);
\draw[ghz example edge,ghzblue] (3.270000,-5.570000) -- (4.510000,-5.570000);
\node[ghz small vertex] at (3.270000,-4.330000) {$1$};
\node[ghz small vertex] at (4.510000,-4.330000) {$2$};
\node[ghz small vertex] at (3.270000,-5.570000) {$3$};
\node[ghz small vertex] at (4.510000,-5.570000) {$4$};
\draw[ghz example edge,ghzred] (5.930000,-4.330000) -- (5.930000,-5.570000);
\draw[ghz example edge,ghzred] (7.170000,-4.330000) -- (7.170000,-5.570000);
\node[ghz small vertex] at (5.930000,-4.330000) {$1$};
\node[ghz small vertex] at (7.170000,-4.330000) {$2$};
\node[ghz small vertex] at (5.930000,-5.570000) {$3$};
\node[ghz small vertex] at (7.170000,-5.570000) {$4$};

\draw[-{Stealth[length=2mm]},ghzink,line width=0.7pt] (1.96,-4.95) -- (2.95,-4.95);
\node[ghz text,font=\fontsize{16}{18}\selectfont] at (5.22,-4.95) {$+$};
\node[ghz small text] at (1.05,-5.98) {Graph $G$};
\node[ghz text,text=ghzblue] at (3.89,-5.98) {$\ket{0000}$};
\node[ghz text,text=ghzred] at (6.55,-5.98) {$\ket{1111}$};
\node[ghz text,font=\fontsize{11}{13}\selectfont] at (3.80,-6.79)
  {$\displaystyle\ket{\mathrm{GHZ}_{4,2}}=\frac{\ket{0000}+\ket{1111}}{\sqrt2}$};
\node[ghz small text,align=center] at (3.80,-7.68)
  {One photon per path selects the two perfect matchings.\\
   Edge weights are unity in this example.};

\node[ghz header] at (8.46,0) {(c)\quad Complete weighted coloured graph};
\node[ghz text] at (12.96,-0.48) {$N=4,\quad D=4$};
\begin{scope}[shift={(12.96,-4.75)}]
\GHZhalfedge{0.000000,3.600000}{0.534771,2.708750}{1.061964,1.813125}{1.581579,0.913125}{2.101194,0.013125}{2.613232,-0.891250}{3.117691,-1.800000}{ghzblue}{ghzblue}
\GHZhalfedge{0.000000,3.600000}{0.571577,2.730000}{1.117173,1.845000}{1.636788,0.945000}{2.156403,0.045000}{2.650038,-0.870000}{3.117691,-1.800000}{ghzred}{ghzred}
\GHZhalfedge{0.000000,3.600000}{0.608383,2.751250}{1.172382,1.876875}{1.691997,0.976875}{2.211612,0.076875}{2.686844,-0.848750}{3.117691,-1.800000}{ghzgreen}{ghzgreen}
\GHZhalfedge{0.000000,3.600000}{0.645189,2.772500}{1.227591,1.908750}{1.747206,1.008750}{2.266821,0.108750}{2.723650,-0.827500}{3.117691,-1.800000}{ghzorange}{ghzorange}
\GHZhalfedge{0.000000,3.600000}{0.681995,2.793750}{1.282800,1.940625}{1.802415,1.040625}{2.322031,0.140625}{2.760456,-0.806250}{3.117691,-1.800000}{ghzblue}{ghzred}
\GHZhalfedge{0.000000,3.600000}{0.718801,2.815000}{1.338009,1.972500}{1.857624,1.072500}{2.377240,0.172500}{2.797262,-0.785000}{3.117691,-1.800000}{ghzred}{ghzgreen}
\GHZhalfedge{0.000000,3.600000}{0.755607,2.836250}{1.393218,2.004375}{1.912834,1.104375}{2.432449,0.204375}{2.834068,-0.763750}{3.117691,-1.800000}{ghzgreen}{ghzorange}
\GHZhalfedge{0.000000,3.600000}{0.792413,2.857500}{1.448427,2.036250}{1.968043,1.136250}{2.487658,0.236250}{2.870874,-0.742500}{3.117691,-1.800000}{ghzorange}{ghzblue}
\GHZhalfedge{0.000000,3.600000}{0.829219,2.878750}{1.503637,2.068125}{2.023252,1.168125}{2.542867,0.268125}{2.907680,-0.721250}{3.117691,-1.800000}{ghzblue}{ghzgreen}
\GHZhalfedge{0.000000,3.600000}{0.866025,2.900000}{1.558846,2.100000}{2.078461,1.200000}{2.598076,0.300000}{2.944486,-0.700000}{3.117691,-1.800000}{ghzred}{ghzorange}
\GHZhalfedge{0.000000,3.600000}{0.902831,2.921250}{1.614055,2.131875}{2.133670,1.231875}{2.653285,0.331875}{2.981292,-0.678750}{3.117691,-1.800000}{ghzgreen}{ghzblue}
\GHZhalfedge{0.000000,3.600000}{0.939638,2.942500}{1.669264,2.163750}{2.188879,1.263750}{2.708494,0.363750}{3.018099,-0.657500}{3.117691,-1.800000}{ghzorange}{ghzred}
\GHZhalfedge{0.000000,3.600000}{0.976444,2.963750}{1.724473,2.195625}{2.244088,1.295625}{2.763704,0.395625}{3.054905,-0.636250}{3.117691,-1.800000}{ghzblue}{ghzorange}
\GHZhalfedge{0.000000,3.600000}{1.013250,2.985000}{1.779682,2.227500}{2.299297,1.327500}{2.818913,0.427500}{3.091711,-0.615000}{3.117691,-1.800000}{ghzred}{ghzblue}
\GHZhalfedge{0.000000,3.600000}{1.050056,3.006250}{1.834891,2.259375}{2.354507,1.359375}{2.874122,0.459375}{3.128517,-0.593750}{3.117691,-1.800000}{ghzgreen}{ghzred}
\GHZhalfedge{0.000000,3.600000}{1.086862,3.027500}{1.890100,2.291250}{2.409716,1.391250}{2.929331,0.491250}{3.165323,-0.572500}{3.117691,-1.800000}{ghzorange}{ghzgreen}
\GHZhalfedge{0.000000,3.600000}{-0.318750,3.000000}{-0.478125,2.400000}{-0.478125,1.800000}{-0.478125,1.200000}{-0.318750,0.600000}{0.000000,0.000000}{ghzblue}{ghzblue}
\GHZhalfedge{0.000000,3.600000}{-0.276250,3.000000}{-0.414375,2.400000}{-0.414375,1.800000}{-0.414375,1.200000}{-0.276250,0.600000}{0.000000,0.000000}{ghzred}{ghzred}
\GHZhalfedge{0.000000,3.600000}{-0.233750,3.000000}{-0.350625,2.400000}{-0.350625,1.800000}{-0.350625,1.200000}{-0.233750,0.600000}{0.000000,0.000000}{ghzgreen}{ghzgreen}
\GHZhalfedge{0.000000,3.600000}{-0.191250,3.000000}{-0.286875,2.400000}{-0.286875,1.800000}{-0.286875,1.200000}{-0.191250,0.600000}{0.000000,0.000000}{ghzorange}{ghzorange}
\GHZhalfedge{0.000000,3.600000}{-0.148750,3.000000}{-0.223125,2.400000}{-0.223125,1.800000}{-0.223125,1.200000}{-0.148750,0.600000}{0.000000,0.000000}{ghzblue}{ghzred}
\GHZhalfedge{0.000000,3.600000}{-0.106250,3.000000}{-0.159375,2.400000}{-0.159375,1.800000}{-0.159375,1.200000}{-0.106250,0.600000}{0.000000,0.000000}{ghzred}{ghzgreen}
\GHZhalfedge{0.000000,3.600000}{-0.063750,3.000000}{-0.095625,2.400000}{-0.095625,1.800000}{-0.095625,1.200000}{-0.063750,0.600000}{0.000000,0.000000}{ghzgreen}{ghzorange}
\GHZhalfedge{0.000000,3.600000}{-0.021250,3.000000}{-0.031875,2.400000}{-0.031875,1.800000}{-0.031875,1.200000}{-0.021250,0.600000}{0.000000,0.000000}{ghzorange}{ghzblue}
\GHZhalfedge{0.000000,3.600000}{0.021250,3.000000}{0.031875,2.400000}{0.031875,1.800000}{0.031875,1.200000}{0.021250,0.600000}{0.000000,0.000000}{ghzblue}{ghzgreen}
\GHZhalfedge{0.000000,3.600000}{0.063750,3.000000}{0.095625,2.400000}{0.095625,1.800000}{0.095625,1.200000}{0.063750,0.600000}{0.000000,0.000000}{ghzred}{ghzorange}
\GHZhalfedge{0.000000,3.600000}{0.106250,3.000000}{0.159375,2.400000}{0.159375,1.800000}{0.159375,1.200000}{0.106250,0.600000}{0.000000,0.000000}{ghzgreen}{ghzblue}
\GHZhalfedge{0.000000,3.600000}{0.148750,3.000000}{0.223125,2.400000}{0.223125,1.800000}{0.223125,1.200000}{0.148750,0.600000}{0.000000,0.000000}{ghzorange}{ghzred}
\GHZhalfedge{0.000000,3.600000}{0.191250,3.000000}{0.286875,2.400000}{0.286875,1.800000}{0.286875,1.200000}{0.191250,0.600000}{0.000000,0.000000}{ghzblue}{ghzorange}
\GHZhalfedge{0.000000,3.600000}{0.233750,3.000000}{0.350625,2.400000}{0.350625,1.800000}{0.350625,1.200000}{0.233750,0.600000}{0.000000,0.000000}{ghzred}{ghzblue}
\GHZhalfedge{0.000000,3.600000}{0.276250,3.000000}{0.414375,2.400000}{0.414375,1.800000}{0.414375,1.200000}{0.276250,0.600000}{0.000000,0.000000}{ghzgreen}{ghzred}
\GHZhalfedge{0.000000,3.600000}{0.318750,3.000000}{0.478125,2.400000}{0.478125,1.800000}{0.478125,1.200000}{0.318750,0.600000}{0.000000,0.000000}{ghzorange}{ghzgreen}
\GHZhalfedge{0.000000,3.600000}{-0.534771,2.708750}{-1.061964,1.813125}{-1.581579,0.913125}{-2.101194,0.013125}{-2.613232,-0.891250}{-3.117691,-1.800000}{ghzblue}{ghzblue}
\GHZhalfedge{0.000000,3.600000}{-0.571577,2.730000}{-1.117173,1.845000}{-1.636788,0.945000}{-2.156403,0.045000}{-2.650038,-0.870000}{-3.117691,-1.800000}{ghzred}{ghzred}
\GHZhalfedge{0.000000,3.600000}{-0.608383,2.751250}{-1.172382,1.876875}{-1.691997,0.976875}{-2.211612,0.076875}{-2.686844,-0.848750}{-3.117691,-1.800000}{ghzgreen}{ghzgreen}
\GHZhalfedge{0.000000,3.600000}{-0.645189,2.772500}{-1.227591,1.908750}{-1.747206,1.008750}{-2.266821,0.108750}{-2.723650,-0.827500}{-3.117691,-1.800000}{ghzorange}{ghzorange}
\GHZhalfedge{0.000000,3.600000}{-0.681995,2.793750}{-1.282800,1.940625}{-1.802415,1.040625}{-2.322031,0.140625}{-2.760456,-0.806250}{-3.117691,-1.800000}{ghzblue}{ghzred}
\GHZhalfedge{0.000000,3.600000}{-0.718801,2.815000}{-1.338009,1.972500}{-1.857624,1.072500}{-2.377240,0.172500}{-2.797262,-0.785000}{-3.117691,-1.800000}{ghzred}{ghzgreen}
\GHZhalfedge{0.000000,3.600000}{-0.755607,2.836250}{-1.393218,2.004375}{-1.912834,1.104375}{-2.432449,0.204375}{-2.834068,-0.763750}{-3.117691,-1.800000}{ghzgreen}{ghzorange}
\GHZhalfedge{0.000000,3.600000}{-0.792413,2.857500}{-1.448427,2.036250}{-1.968043,1.136250}{-2.487658,0.236250}{-2.870874,-0.742500}{-3.117691,-1.800000}{ghzorange}{ghzblue}
\GHZhalfedge{0.000000,3.600000}{-0.829219,2.878750}{-1.503637,2.068125}{-2.023252,1.168125}{-2.542867,0.268125}{-2.907680,-0.721250}{-3.117691,-1.800000}{ghzblue}{ghzgreen}
\GHZhalfedge{0.000000,3.600000}{-0.866025,2.900000}{-1.558846,2.100000}{-2.078461,1.200000}{-2.598076,0.300000}{-2.944486,-0.700000}{-3.117691,-1.800000}{ghzred}{ghzorange}
\GHZhalfedge{0.000000,3.600000}{-0.902831,2.921250}{-1.614055,2.131875}{-2.133670,1.231875}{-2.653285,0.331875}{-2.981292,-0.678750}{-3.117691,-1.800000}{ghzgreen}{ghzblue}
\GHZhalfedge{0.000000,3.600000}{-0.939638,2.942500}{-1.669264,2.163750}{-2.188879,1.263750}{-2.708494,0.363750}{-3.018099,-0.657500}{-3.117691,-1.800000}{ghzorange}{ghzred}
\GHZhalfedge{0.000000,3.600000}{-0.976444,2.963750}{-1.724473,2.195625}{-2.244088,1.295625}{-2.763704,0.395625}{-3.054905,-0.636250}{-3.117691,-1.800000}{ghzblue}{ghzorange}
\GHZhalfedge{0.000000,3.600000}{-1.013250,2.985000}{-1.779682,2.227500}{-2.299297,1.327500}{-2.818913,0.427500}{-3.091711,-0.615000}{-3.117691,-1.800000}{ghzred}{ghzblue}
\GHZhalfedge{0.000000,3.600000}{-1.050056,3.006250}{-1.834891,2.259375}{-2.354507,1.359375}{-2.874122,0.459375}{-3.128517,-0.593750}{-3.117691,-1.800000}{ghzgreen}{ghzred}
\GHZhalfedge{0.000000,3.600000}{-1.086862,3.027500}{-1.890100,2.291250}{-2.409716,1.391250}{-2.929331,0.491250}{-3.165323,-0.572500}{-3.117691,-1.800000}{ghzorange}{ghzgreen}
\GHZhalfedge{3.117691,-1.800000}{2.757451,-1.223954}{2.317523,-0.785932}{1.797908,-0.485932}{1.278293,-0.185932}{0.678990,-0.023954}{0.000000,0.000000}{ghzblue}{ghzblue}
\GHZhalfedge{3.117691,-1.800000}{2.736201,-1.260760}{2.285648,-0.841141}{1.766033,-0.541141}{1.246418,-0.241141}{0.657740,-0.060760}{0.000000,0.000000}{ghzred}{ghzred}
\GHZhalfedge{3.117691,-1.800000}{2.714951,-1.297567}{2.253773,-0.896350}{1.734158,-0.596350}{1.214543,-0.296350}{0.636490,-0.097567}{0.000000,0.000000}{ghzgreen}{ghzgreen}
\GHZhalfedge{3.117691,-1.800000}{2.693701,-1.334373}{2.221898,-0.951559}{1.702283,-0.651559}{1.182668,-0.351559}{0.615240,-0.134373}{0.000000,0.000000}{ghzorange}{ghzorange}
\GHZhalfedge{3.117691,-1.800000}{2.672451,-1.371179}{2.190023,-1.006768}{1.670408,-0.706768}{1.150793,-0.406768}{0.593990,-0.171179}{0.000000,0.000000}{ghzblue}{ghzred}
\GHZhalfedge{3.117691,-1.800000}{2.651201,-1.407985}{2.158148,-1.061977}{1.638533,-0.761977}{1.118918,-0.461977}{0.572740,-0.207985}{0.000000,0.000000}{ghzred}{ghzgreen}
\GHZhalfedge{3.117691,-1.800000}{2.629951,-1.444791}{2.126273,-1.117186}{1.606658,-0.817186}{1.087043,-0.517186}{0.551490,-0.244791}{0.000000,0.000000}{ghzgreen}{ghzorange}
\GHZhalfedge{3.117691,-1.800000}{2.608701,-1.481597}{2.094398,-1.172395}{1.574783,-0.872395}{1.055168,-0.572395}{0.530240,-0.281597}{0.000000,0.000000}{ghzorange}{ghzblue}
\GHZhalfedge{3.117691,-1.800000}{2.587451,-1.518403}{2.062523,-1.227605}{1.542908,-0.927605}{1.023293,-0.627605}{0.508990,-0.318403}{0.000000,0.000000}{ghzblue}{ghzgreen}
\GHZhalfedge{3.117691,-1.800000}{2.566201,-1.555209}{2.030648,-1.282814}{1.511033,-0.982814}{0.991418,-0.682814}{0.487740,-0.355209}{0.000000,0.000000}{ghzred}{ghzorange}
\GHZhalfedge{3.117691,-1.800000}{2.544951,-1.592015}{1.998773,-1.338023}{1.479158,-1.038023}{0.959543,-0.738023}{0.466490,-0.392015}{0.000000,0.000000}{ghzgreen}{ghzblue}
\GHZhalfedge{3.117691,-1.800000}{2.523701,-1.628821}{1.966898,-1.393232}{1.447283,-1.093232}{0.927668,-0.793232}{0.445240,-0.428821}{0.000000,0.000000}{ghzorange}{ghzred}
\GHZhalfedge{3.117691,-1.800000}{2.502451,-1.665627}{1.935023,-1.448441}{1.415408,-1.148441}{0.895793,-0.848441}{0.423990,-0.465627}{0.000000,0.000000}{ghzblue}{ghzorange}
\GHZhalfedge{3.117691,-1.800000}{2.481201,-1.702433}{1.903148,-1.503650}{1.383533,-1.203650}{0.863918,-0.903650}{0.402740,-0.502433}{0.000000,0.000000}{ghzred}{ghzblue}
\GHZhalfedge{3.117691,-1.800000}{2.459951,-1.739240}{1.871273,-1.558859}{1.351658,-1.258859}{0.832043,-0.958859}{0.381490,-0.539240}{0.000000,0.000000}{ghzgreen}{ghzred}
\GHZhalfedge{3.117691,-1.800000}{2.438701,-1.776046}{1.839398,-1.614068}{1.319783,-1.314068}{0.800168,-1.014068}{0.360240,-0.576046}{0.000000,0.000000}{ghzorange}{ghzgreen}
\GHZhalfedge{3.117691,-1.800000}{2.078461,-1.817500}{1.039230,-1.826250}{0.000000,-1.826250}{-1.039230,-1.826250}{-2.078461,-1.817500}{-3.117691,-1.800000}{ghzblue}{ghzblue}
\GHZhalfedge{3.117691,-1.800000}{2.078461,-1.860000}{1.039230,-1.890000}{0.000000,-1.890000}{-1.039230,-1.890000}{-2.078461,-1.860000}{-3.117691,-1.800000}{ghzred}{ghzred}
\GHZhalfedge{3.117691,-1.800000}{2.078461,-1.902500}{1.039230,-1.953750}{0.000000,-1.953750}{-1.039230,-1.953750}{-2.078461,-1.902500}{-3.117691,-1.800000}{ghzgreen}{ghzgreen}
\GHZhalfedge{3.117691,-1.800000}{2.078461,-1.945000}{1.039230,-2.017500}{0.000000,-2.017500}{-1.039230,-2.017500}{-2.078461,-1.945000}{-3.117691,-1.800000}{ghzorange}{ghzorange}
\GHZhalfedge{3.117691,-1.800000}{2.078461,-1.987500}{1.039230,-2.081250}{0.000000,-2.081250}{-1.039230,-2.081250}{-2.078461,-1.987500}{-3.117691,-1.800000}{ghzblue}{ghzred}
\GHZhalfedge{3.117691,-1.800000}{2.078461,-2.030000}{1.039230,-2.145000}{0.000000,-2.145000}{-1.039230,-2.145000}{-2.078461,-2.030000}{-3.117691,-1.800000}{ghzred}{ghzgreen}
\GHZhalfedge{3.117691,-1.800000}{2.078461,-2.072500}{1.039230,-2.208750}{0.000000,-2.208750}{-1.039230,-2.208750}{-2.078461,-2.072500}{-3.117691,-1.800000}{ghzgreen}{ghzorange}
\GHZhalfedge{3.117691,-1.800000}{2.078461,-2.115000}{1.039230,-2.272500}{0.000000,-2.272500}{-1.039230,-2.272500}{-2.078461,-2.115000}{-3.117691,-1.800000}{ghzorange}{ghzblue}
\GHZhalfedge{3.117691,-1.800000}{2.078461,-2.157500}{1.039230,-2.336250}{0.000000,-2.336250}{-1.039230,-2.336250}{-2.078461,-2.157500}{-3.117691,-1.800000}{ghzblue}{ghzgreen}
\GHZhalfedge{3.117691,-1.800000}{2.078461,-2.200000}{1.039230,-2.400000}{0.000000,-2.400000}{-1.039230,-2.400000}{-2.078461,-2.200000}{-3.117691,-1.800000}{ghzred}{ghzorange}
\GHZhalfedge{3.117691,-1.800000}{2.078461,-2.242500}{1.039230,-2.463750}{0.000000,-2.463750}{-1.039230,-2.463750}{-2.078461,-2.242500}{-3.117691,-1.800000}{ghzgreen}{ghzblue}
\GHZhalfedge{3.117691,-1.800000}{2.078461,-2.285000}{1.039230,-2.527500}{0.000000,-2.527500}{-1.039230,-2.527500}{-2.078461,-2.285000}{-3.117691,-1.800000}{ghzorange}{ghzred}
\GHZhalfedge{3.117691,-1.800000}{2.078461,-2.327500}{1.039230,-2.591250}{0.000000,-2.591250}{-1.039230,-2.591250}{-2.078461,-2.327500}{-3.117691,-1.800000}{ghzblue}{ghzorange}
\GHZhalfedge{3.117691,-1.800000}{2.078461,-2.370000}{1.039230,-2.655000}{0.000000,-2.655000}{-1.039230,-2.655000}{-2.078461,-2.370000}{-3.117691,-1.800000}{ghzred}{ghzblue}
\GHZhalfedge{3.117691,-1.800000}{2.078461,-2.412500}{1.039230,-2.718750}{0.000000,-2.718750}{-1.039230,-2.718750}{-2.078461,-2.412500}{-3.117691,-1.800000}{ghzgreen}{ghzred}
\GHZhalfedge{3.117691,-1.800000}{2.078461,-2.455000}{1.039230,-2.782500}{0.000000,-2.782500}{-1.039230,-2.782500}{-2.078461,-2.455000}{-3.117691,-1.800000}{ghzorange}{ghzgreen}
\GHZhalfedge{0.000000,0.000000}{-0.678990,-0.023954}{-1.278293,-0.185932}{-1.797908,-0.485932}{-2.317523,-0.785932}{-2.757451,-1.223954}{-3.117691,-1.800000}{ghzblue}{ghzblue}
\GHZhalfedge{0.000000,0.000000}{-0.657740,-0.060760}{-1.246418,-0.241141}{-1.766033,-0.541141}{-2.285648,-0.841141}{-2.736201,-1.260760}{-3.117691,-1.800000}{ghzred}{ghzred}
\GHZhalfedge{0.000000,0.000000}{-0.636490,-0.097567}{-1.214543,-0.296350}{-1.734158,-0.596350}{-2.253773,-0.896350}{-2.714951,-1.297567}{-3.117691,-1.800000}{ghzgreen}{ghzgreen}
\GHZhalfedge{0.000000,0.000000}{-0.615240,-0.134373}{-1.182668,-0.351559}{-1.702283,-0.651559}{-2.221898,-0.951559}{-2.693701,-1.334373}{-3.117691,-1.800000}{ghzorange}{ghzorange}
\GHZhalfedge{0.000000,0.000000}{-0.593990,-0.171179}{-1.150793,-0.406768}{-1.670408,-0.706768}{-2.190023,-1.006768}{-2.672451,-1.371179}{-3.117691,-1.800000}{ghzblue}{ghzred}
\GHZhalfedge{0.000000,0.000000}{-0.572740,-0.207985}{-1.118918,-0.461977}{-1.638533,-0.761977}{-2.158148,-1.061977}{-2.651201,-1.407985}{-3.117691,-1.800000}{ghzred}{ghzgreen}
\GHZhalfedge{0.000000,0.000000}{-0.551490,-0.244791}{-1.087043,-0.517186}{-1.606658,-0.817186}{-2.126273,-1.117186}{-2.629951,-1.444791}{-3.117691,-1.800000}{ghzgreen}{ghzorange}
\GHZhalfedge{0.000000,0.000000}{-0.530240,-0.281597}{-1.055168,-0.572395}{-1.574783,-0.872395}{-2.094398,-1.172395}{-2.608701,-1.481597}{-3.117691,-1.800000}{ghzorange}{ghzblue}
\GHZhalfedge{0.000000,0.000000}{-0.508990,-0.318403}{-1.023293,-0.627605}{-1.542908,-0.927605}{-2.062523,-1.227605}{-2.587451,-1.518403}{-3.117691,-1.800000}{ghzblue}{ghzgreen}
\GHZhalfedge{0.000000,0.000000}{-0.487740,-0.355209}{-0.991418,-0.682814}{-1.511033,-0.982814}{-2.030648,-1.282814}{-2.566201,-1.555209}{-3.117691,-1.800000}{ghzred}{ghzorange}
\GHZhalfedge{0.000000,0.000000}{-0.466490,-0.392015}{-0.959543,-0.738023}{-1.479158,-1.038023}{-1.998773,-1.338023}{-2.544951,-1.592015}{-3.117691,-1.800000}{ghzgreen}{ghzblue}
\GHZhalfedge{0.000000,0.000000}{-0.445240,-0.428821}{-0.927668,-0.793232}{-1.447283,-1.093232}{-1.966898,-1.393232}{-2.523701,-1.628821}{-3.117691,-1.800000}{ghzorange}{ghzred}
\GHZhalfedge{0.000000,0.000000}{-0.423990,-0.465627}{-0.895793,-0.848441}{-1.415408,-1.148441}{-1.935023,-1.448441}{-2.502451,-1.665627}{-3.117691,-1.800000}{ghzblue}{ghzorange}
\GHZhalfedge{0.000000,0.000000}{-0.402740,-0.502433}{-0.863918,-0.903650}{-1.383533,-1.203650}{-1.903148,-1.503650}{-2.481201,-1.702433}{-3.117691,-1.800000}{ghzred}{ghzblue}
\GHZhalfedge{0.000000,0.000000}{-0.381490,-0.539240}{-0.832043,-0.958859}{-1.351658,-1.258859}{-1.871273,-1.558859}{-2.459951,-1.739240}{-3.117691,-1.800000}{ghzgreen}{ghzred}
\GHZhalfedge{0.000000,0.000000}{-0.360240,-0.576046}{-0.800168,-1.014068}{-1.319783,-1.314068}{-1.839398,-1.614068}{-2.438701,-1.776046}{-3.117691,-1.800000}{ghzorange}{ghzgreen}
\node[ghz vertex] at (0.000000,3.600000) {$1$};
\node[ghz vertex] at (3.117691,-1.800000) {$2$};
\node[ghz vertex] at (0.000000,0.000000) {$3$};
\node[ghz vertex] at (-3.117691,-1.800000) {$4$};

\end{scope}
\node[ghz text] at (12.96,-8.02)
  {$\binom42 D^2=6\times16=96\ \text{edges}$};
\end{tikzpicture}

\endgroup
}
\caption{
Graph-based description of photonic experiments.
(a) Vertices denote photonic paths; an edge represents a photon-pair.
The colour of the edge encode the internal modes of the photons at their respective
endpoints: blue, red, green, and orange denote modes $0$, $1$, $2$, and $3$.
The complex edge weight $w_e$ specifies the amplitude and phase.
(b) For the illustrated four-vertex graph, all displayed edge weights equal one.
Conditioning on one photon per path leads to two perfect matchings, whose coherent
contributions give $\ket{\mathrm{GHZ}_{4,2}}=(\ket{0000}+\ket{1111})/\sqrt2$.
(c) The unrestricted $N=D=4$ candidate graph contains all $D^2=16$ endpoint-colour
combinations for each of its six vertex pairs, giving $96$ edges in total:
$24$ monochromatic and $72$ bichromatic.
Each edge corresponds to one finite complex weight $W_{rs}^{ab}$
[Eq.~\eqref{eq:effective-edge-weight}], including the possibility of zero weight. Due to our theorem, we know that this graph cannot lead to a 4-dimensional GHZ state.
}
\label{fig:graph_overview}
\end{figure*}
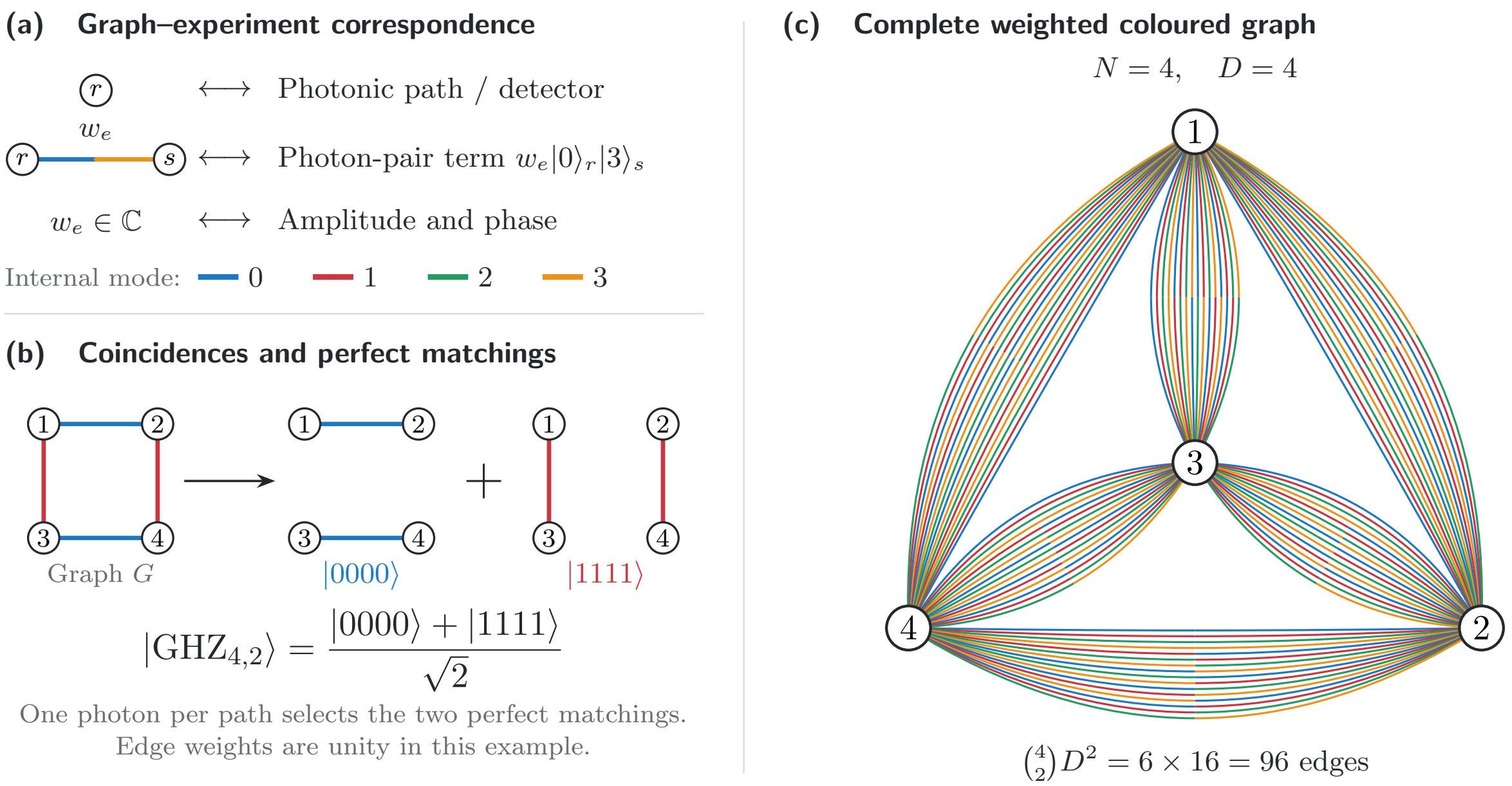

For a colouring \(\mathbf c\in C_D^N\), define the corresponding computational-basis state by \(\ket{\mathbf c}:=\ket{c_{v_1},\ldots,c_{v_N}}\). The conditioned quantum state associated with the graph \(G\) is, up to normalisation,
\begin{equation}
\ket{\psi_G}
=
\sum_{\mathbf c\in C_D^N}
\omega_{\mathbf c}
\ket{\mathbf c}.
\label{eq:conditioned-graph-state}
\end{equation}
To realise the GHZ target in Eq.~\eqref{eq:ghz-state}, only the fully monochromatic basis states \(\ket{j}^{\otimes N}\), with \(j\in C_D\), may have nonzero amplitude; every other colouring must be suppressed. Figure~\ref{fig:graph_overview} summarises the graph--experiment correspondence, a simple GHZ-generating graph, and the complete colour-resolved four-vertex multigraph considered in the no-go analysis.

\begin{table*}[!t]
\centering
\small
\setlength{\tabcolsep}{4pt}
\caption{Current state of results on monochromatic quantum graphs.
Here \(N\) is even, and \(D_{\max}\) denotes the maximum achievable
dimension under the stated graph and weight constraints.
The 2017 conjectured bounds are distinguished from subsequent
proved results.}
\label{tab:state-of-the-art}

\begin{tabular}{|p{0.17\textwidth}|p{0.22\textwidth}|p{0.20\textwidth}|p{0.30\textwidth}|}
\hline
\textbf{Date} &
\textbf{Graph Restrictions} &
\textbf{Bound} &
\textbf{Comment and Refs.} \\
\hline

2017 &
-- &
$N=4$: $D_{\max}=3$;\newline
$N\geq6$: $D_{\max}=2$ &
Monochromatic Quantum Graph conjecture
~\cite{krenn2017quantum,krenn2017mathoverflow,krenn2019questions}. \\
\hline

2017 &
$w_e\in\mathbb{R}_{\geq0}$ &
$N=4$: $D_{\max}=3$;\newline
$N\geq6$: $D_{\max}=2$ &
Bogdanov's lemma~\cite{bogdanov2017mathoverflow}. \\
\hline

2022 &
No bichromatic edges &
$N=6$: $D_{\max}=2$;\newline
$N=8$: $D_{\max}\leq3$ &
SAT-solver constraint~\cite{cervera2022design}. \\
\hline

2023 &
Unrestricted &
$N=4$: $D_{\max}=3$ &
Gr\"{o}bner basis constraint
~\cite{mantey2023graphn4solution}. \\
\hline

2024 &
Vertex connectivity $\kappa(G)\leq2$, or cubic graphs &
$N\geq6$: $D_{\max}=2$ &
Combinatorial proof~\cite{chandran2024krenngu}. \\
\hline

2024 &
No multi-edges &
$N\geq6$:
$D_{\max}<N/\sqrt{2}$ &
Combinatorial proof
~\cite{chandran2024graphtheoretic}. \\
\hline

\textbf{April 2026 (This work)} &
\textbf{Unrestricted} &
\textbf{$N\geq4$: $D_{\max}<N$} &
\textbf{AI-discovered with \textit{AlphaProof Nexus};
formally verified, human-digested (this manuscript);
tensor-algebraic.} \\
\hline

\textit{July 2026} &
\textit{Unrestricted} &
\textit{$N\geq6$: $D_{\max}\leq N-2$} &
\textit{Kenta Kitamura; AI-assisted with \textit{ChatGPT/Codex
GPT-5.6 Sol}; formally verified. Not human-digested.
~\cite{kitamura2026sharpbound}.} \\
\hline

\textit{July 2026} &
\textit{$w_e\in\mathbb{Z}$} &
\textit{$N\geq6$: $D_{\max}=2$} &
\textit{Kenta Kitamura; AI-assisted with \textit{ChatGPT/Codex
GPT-5.6 Sol}; formally verified. Not human-digested.
~\cite{kitamura2026integers}.} \\
\hline

\textit{September 2026} &
\textit{Unrestricted} &
\textit{$N\geq6$: $D_{\max}=2$} &
\textit{Rishi Rajasekaran; AI-assisted with
Codex/Claude; formally verified.
Not human-digested.
\cite{rajasekaran2026krenngu}} \\
\hline
\end{tabular}
\end{table*}

\subsection*{The conjecture and previous progress}

In the graph-theoretic representation, a \(D\)-dimensional \(N\)-photon GHZ state has nonzero
amplitudes only on fully monochromatic inherited vertex
colourings. For every colouring \(\mathbf c\in C_D^N\), define the GHZ target amplitude tensor
\(\Delta=(\Delta_{\mathbf c})_{\mathbf c\in C_D^N}\) componentwise by
\begin{equation}
\Delta_{\mathbf c}
:=
\begin{cases}
1, & \text{if } \mathbf c=(j,j,\ldots,j)\\
   & \text{for some } j\in C_D,\\[2mm]
0, & \text{otherwise}.
\end{cases}
\label{eq:monochromatic-target-amplitudes}
\end{equation}
For simplicity, \(\Delta\) is left unnormalised: normalising it gives the GHZ state in Eq.~\eqref{eq:ghz-state}. Write \(\omega=(\omega_{\mathbf c})_{\mathbf c\in C_D^N}\) for the graph amplitude tensor defined componentwise in Eq.~\eqref{eq:coloring-amplitude}. We call a graph a \textit{monochromatic quantum graph} when \(\omega=\Delta\). Setting the common nonzero monochromatic amplitude to \(1\) entails no loss of generality, since it can be absorbed by uniformly rescaling the edge weights.\footnote{A graph physically realises the GHZ target when \(\omega=\alpha\Delta\) for some \(\alpha\in\mathbb C\setminus\{0\}\). For even \(N\), every perfect matching contains \(N/2\) edges. Rescaling all edge weights to \(w'_e=\beta w_e\), where \(\beta^{N/2}=\alpha^{-1}\), therefore gives \(\omega'=\beta^{N/2}\omega=\Delta\). Such a \(\beta\) always exists over \(\mathbb C\).}

This leads to a natural question:
\begin{question}
For a given even number of vertices \(N\), what is the largest possible \(D\)
for which a monochromatic quantum graph can exist?
\end{question}

Significant numerical results and theoretical arguments have led to the following conjecture:

\begin{conjecture}
For a given even number of vertices $N$, a monochromatic quantum graph can only exist for
$N=4, D\leq3$ and $N\geq6, D=2$.
\end{conjecture}

Table~\ref{tab:state-of-the-art} records the detailed partial results. For unrestricted complex-weighted multigraphs, the conjectured upper bound is known for \(N=4\), but for even \(N\geq6\) the intermediate range \(3\leq D<N\) remains open. The theorem below excludes the complementary regime \(D\geq N\).

\section*{The origin of the proof}

The proof presented here was discovered in the context of the \textit{Formal Conjectures} project \cite{FormalConjectures, FormalConjecturesPaper}, an open repository of mathematical conjectures formalised in Lean~\cite{demoura2021lean}. Lean is an interactive theorem prover in which definitions, theorem statements, and proofs are written as machine-checkable code. Mathlib is a mathematical library available in Lean, providing a large shared foundation of formalised mathematics on which new formal proofs can build~\cite{mathlib2020}. The Formal Conjectures repository collects conjectures whose statements have been translated into Lean, thereby turning open problems into targets for automated and AI-assisted theorem proving, as well as a mechanism to canonise statements of open conjectures. At the time of writing, the project lists 1979 formalised problems, of which 1090 are open and 889 are marked solved.

The formal proof was found by \textit{AlphaProof Nexus}, an AI-driven formal proof search framework that combines frontier language models with feedback from the Lean compiler~\cite{tsoukalas2026advancing}. In this framework, the input is a Lean file containing definitions and a target theorem whose proof is missing; the agent iteratively edits the proof, receives compiler feedback, introduces auxiliary lemmas, and searches for a proof that compiles without unresolved goals. In the present case, the formal proof led to the tensor-algebraic argument developed in the next section.

The monochromatic quantum graph conjecture was formalised in Lean as part
of the \textit{Formal Conjectures} project
\cite{FormalConjectures}. The formalisation encodes both
the complex-weighted graphs and the coherent sums over
their perfect matchings. In Lean, the vertices are represented
by \texttt{Fin N}, while \texttt{EdgeN N D} specifies two
vertices together with their endpoint colours. The
edge-weight function is defined as

\noindent\begin{minipage}{\columnwidth}
\begin{lstlisting}[style=leanpaper]
/-- Weights on edges. -/
abbrev WeightsN (N D : Nat) (α : Type) :=
  EdgeN N D → α
\end{lstlisting}
\end{minipage}

The sum over perfect matchings is computed recursively.
The function \texttt{pmSumListAux} selects the first
remaining vertex, pairs it with each possible partner,
multiplies the corresponding edge weight by the contribution
from the remaining vertices, and sums over all choices.
This implements the coherent addition of the amplitudes
of all perfect matchings:

\noindent\begin{minipage}{\columnwidth}
\begin{lstlisting}[style=leanpaper]
def pmSumListAux {α : Type} [Semiring α] {N D : Nat}
    (W : WeightsN N D α) (ι : V N → Fin D) :
    Nat → List (V N) → α
  | 0, _ => 1
  | 1, _ => 0
  | _ + 2, [] => 1
  | _ + 2, [_] => 0
  | n + 2, v :: vs =>
      (vs.map (fun u =>
          W (mkEdge v u (ι v) (ι u)) *
            pmSumListAux W ι n (vs.erase u)
        )).sum
\end{lstlisting}
\end{minipage}

Two further definitions apply this recursion to a
vertex list and ultimately to the full graph, using the
canonical ordering of its \(N\) vertices:

\noindent\begin{minipage}{\columnwidth}
\begin{lstlisting}[style=leanpaper]
/-- Perfect-matching sum on a list. -/
def pmSumList {α : Type} [Semiring α] {N D : Nat}
    (W : WeightsN N D α)
    (ι : V N → Fin D) (L : List (V N)) : α :=
  pmSumListAux W ι L.length L

/-- Perfect-matching sum for K_N. -/
def pmSumN {α : Type} [Semiring α] (N D : Nat)
    (W : WeightsN N D α) (ι : V N → Fin D) : α :=
  pmSumList W ι (vertices N)
\end{lstlisting}
\end{minipage}

The central predicate \texttt{EqSystemN} then defines
the conditions for a monochromatic quantum graph.
For every assignment of colours
\(\iota:V_N\to C_D\), the sum over perfect matchings
must equal \(1\) if all vertices inherit the same colour,
and \(0\) otherwise. In Lean, these conditions read

\noindent\begin{minipage}{\columnwidth}
\begin{lstlisting}[style=leanpaper]
def EqSystemN {α : Type} [Semiring α]
    (N D : Nat) (W : WeightsN N D α) : Prop :=
  ∀ ι : V N → Fin D,
    pmSumN N D W ι =
      (if allEqual ι then (1 : α) else (0 : α))
\end{lstlisting}
\end{minipage}

Here, \texttt{allEqual} tests whether the inherited
vertex colouring is monochromatic. The conjecture for
even \(N\geq6\) and \(D\geq3\) can therefore be
expressed compactly as follows (with the proof body
omitted):

\noindent\begin{minipage}{\columnwidth}
\begin{lstlisting}[style=leanpaper]
theorem eqSystem_no_solution_ge6_ge3 :
    ∀ N D : Nat, N ≥ 6 → Even N → D ≥ 3 →
      ¬ ∃ W : WeightsN N D ℂ, EqSystemN N D W
\end{lstlisting}
\end{minipage}

The AI-discovered formal proof establishes a general
non-existence result for all even \(N\geq4\) in the
case \(D=N\), without restrictions on the complex edge
weights. The corresponding Lean theorem states

\noindent\begin{minipage}{\columnwidth}
\begin{lstlisting}[style=leanpaper]
theorem
    eqSystem_no_solution_even_ge4_d_eq_n_explicit :
    ∀ N : Nat, N ≥ 4 → Even N →
      ¬ ∃ W : WeightsN N N ℂ, EqSystemN N N W
\end{lstlisting}
\end{minipage}

The symbols \(\forall\), \(\exists\), and \(\neg\) denote universal quantification, existence, and negation, respectively. Thus, the theorem establishes that no assignment of complex edge weights can satisfy the monochromatic quantum graph equations when \(D=N\). The complete machine-checked proof is available \href{https://github.com/google-deepmind/formal-conjectures/blob/af88acbf9da0f26e3e934743a819e986e02f6875/FormalConjectures/Paper/MonochromaticQuantumGraph.lean#L1006}{online}. As we show below, this result extends to \(D>N\) by restricting the colour set.

\section*{The proof for the non-existence of monochromatic graphs with \texorpdfstring{$D\geq N$}{D >= N}}
We prove the following theorem. The nontrivial case is even \(N\), since
perfect matchings cover vertices in pairs.

\begin{theorem}
Let \(N\geq4\) be even. There is no complex-weighted monochromatic quantum
graph with \(N\) vertices and \(D\geq N\) colours.
\end{theorem}

We first introduce effective edge weights and a weighted sum of amplitudes whose coefficients factor vertex by vertex. We then prove the theorem for \(N=D=4\), using a \(3\times4\) matrix formed from the edges incident to one chosen vertex. This example shows how to make the weighted sum vanish for the graph while keeping a nonzero contribution from the GHZ target. The same construction applies to an \((N-1)\times N\) matrix for arbitrary even \(N\). Finally, the case \(D>N\) follows by restricting to \(N\) colours.

\subsection*{Effective edge weights}

For \(1\le r<s\le N\) and \(a,b\in C_D\), let \(E_{rs}^{ab}\) be the edges joining \(v_r\) and \(v_s\) with endpoint colours \(a\) and \(b\), respectively. Define
\begin{equation}
W_{rs}^{ab}
:=
\sum_{e\in E_{rs}^{ab}}w_e.
\label{eq:effective-edge-weight}
\end{equation}
Empty sums are zero. Summing parallel edges leaves the amplitudes unchanged, giving
\begin{equation}
\omega_{\mathbf c}
=
\sum_{P\in\mathcal{PM}(K_N)}
\prod_{\substack{\{v_r,v_s\}\in P\\ r<s}}
W_{rs}^{c_{v_r}c_{v_s}}.
\label{eq:effective-pairing-amplitude}
\end{equation}
Here \(K_N\) denotes the complete simple graph on the vertex set \(V(G)\), and \(\mathcal{PM}(K_N)\) denotes its perfect matchings, i.e., all partitions of \(V(G)\) into disjoint pairs.

\subsection*{The separating weighted-sum functional}
The main idea is to separate the graph amplitude tensor \(\omega\) from the target amplitude tensor \(\Delta\) using a carefully chosen functional dependent on the graph. If a monochromatic graph existed, the two tensors would agree entrywise. Consequently, for any functional \(\mathcal F\),
\begin{equation}
\omega=\Delta
\quad\Longrightarrow\quad
\mathcal F(\omega)=\mathcal F(\Delta).
\label{eq:functional-equality}
\end{equation}
Therefore, to prove non-existence, it is enough to construct a functional $\mathcal F(X)$ that takes different values on the two tensors. In the \(D=N\) case, we will construct \(\mathcal F\) so that
\begin{equation}
\mathcal F(\omega)=0
\qquad\text{but}\qquad
\mathcal F(\Delta)\neq0.
\label{eq:separation-condition}
\end{equation}
These different values would contradict the equality required for a monochromatic graph.

We first define this functional for the case \(D=N\), using the square-case colour set
\begin{equation}
C_N:=\{0,\ldots,N-1\}.
\label{eq:general-colour-set}
\end{equation}
For each vertex \(v_r\), choose a coefficient function \(\lambda_r:C_N\to\mathbb{C}\).
The number \(\lambda_r(j)\) is the coefficient assigned to the event that vertex \(v_r\) inherits colour \(j\). Once the functions \(\lambda_1,\ldots,\lambda_N\) have been chosen, define
\begin{equation}
\mathcal F(X)
=
\sum_{\mathbf c\in C_N^N}
\left(
\prod_{r=1}^N \lambda_r(c_{v_r})
\right)
X_{\mathbf c}.
\label{eq:weighted-sum-function}
\end{equation}
Here \(X\) is an amplitude array indexed by vertex colourings \(\mathbf c=(c_{v_1},\ldots,c_{v_N})\). Viewing each \(\lambda_r\) as a vector of colour coefficients, the product vector \(\lambda_1\otimes\cdots\otimes\lambda_N\) has entries \(\prod_{r=1}^N\lambda_r(c_{v_r})\). This factorisation lets each perfect-matching contribution separate into pair contributions.

The target array \(\Delta\) from Eq.~\eqref{eq:monochromatic-target-amplitudes} is nonzero only on monochromatic colourings, so its weighted sum is
\begin{equation}
\mathcal F(\Delta)
=
\sum_{j\in C_N}
\prod_{r=1}^N \lambda_r(j).
\label{eq:delta-weighted-sum}
\end{equation}
For the graph amplitudes, the sum runs over all inherited vertex colourings:
\begin{equation}
\mathcal F(\omega)
=
\sum_{\mathbf c\in C_N^N}
\left(
\prod_{r=1}^N \lambda_r(c_{v_r})
\right)
\omega_{\mathbf c}.
\label{eq:omega-weighted-sum}
\end{equation}
We will choose the coefficient functions \(\lambda_r\) so that the graph sum vanishes while the target sum remains nonzero. This contradicts \(\omega=\Delta\) and rules out the assumed monochromatic graph.

\subsection*{A complete proof for \texorpdfstring{$N=4,D=4$}{N = 4, D = 4}}

We now make the construction explicit by proving the first nontrivial case from beginning to end. Let the four-colour set be
\begin{equation}
C_4:=\{0,1,2,3\}.
\label{eq:n4-colour-set}
\end{equation}
Assume, for contradiction, that a four-vertex, four-colour monochromatic quantum graph exists. Its amplitude array then obeys
\begin{equation}
\begin{aligned}
\omega_{c_1,c_2,c_3,c_4}
&=\Delta_{c_1,c_2,c_3,c_4},\\
&\text{for every }(c_1,c_2,c_3,c_4)\in C_4^4.
\end{aligned}
\label{eq:n4-assumed-table-equality}
\end{equation}
We will construct one functional \(\mathcal F\) for which the graph array and the GHZ target array give different values.

\paragraph*{Step 1: Write the four-vertex amplitude explicitly.}
For an inherited vertex colouring \((c_1,c_2,c_3,c_4)\), the four-vertex amplitude is
\begin{equation}
\omega_{c_1,c_2,c_3,c_4}
= W_{12}^{c_1c_2}W_{34}^{c_3c_4}
+ W_{13}^{c_1c_3}W_{24}^{c_2c_4}
+ W_{14}^{c_1c_4}W_{23}^{c_2c_3}.
\label{eq:four-vertex-effective-amplitude-improved}
\end{equation}
The three terms correspond to the three perfect matchings.
For \(1\le r<s\le4\), define the contracted edge amplitude
\begin{equation}
B_{rs}
=
\sum_{x,y\in C_4}
\lambda_r(x)\lambda_s(y)W_{rs}^{xy}.
\label{eq:B-rs-four-vertex}
\end{equation}
When we apply the functional in Eq.~\eqref{eq:weighted-sum-function} to the four-vertex amplitude, the product coefficient
\(\lambda_1(c_1)\lambda_2(c_2)\lambda_3(c_3)\lambda_4(c_4)\) separates the colour sums over the two pairs of each perfect matching. For example, the contribution of the first term in Eq.~\eqref{eq:four-vertex-effective-amplitude-improved} is
\begin{align}
&\sum_{c_1,c_2,c_3,c_4\in C_4}
\lambda_1(c_1)\lambda_2(c_2)\lambda_3(c_3)\lambda_4(c_4)
W_{12}^{c_1c_2}W_{34}^{c_3c_4}
\nonumber\\
&\quad=
\left(\sum_{c_1,c_2\in C_4}
\lambda_1(c_1)\lambda_2(c_2)W_{12}^{c_1c_2}\right)
\nonumber\\[-0.2em]
&\qquad\times
\left(\sum_{c_3,c_4\in C_4}
\lambda_3(c_3)\lambda_4(c_4)W_{34}^{c_3c_4}\right)
\nonumber\\
&\quad=B_{12}B_{34}.
\label{eq:first-matching-factorization-improved}
\end{align}
The other two matching contributions factor in the same way, giving
\begin{equation}
\boxed{
\mathcal F(\omega)
=
B_{12}B_{34}+B_{13}B_{24}+B_{14}B_{23}.
}
\label{eq:Fomega-four-explicit}
\end{equation}
\paragraph*{Step 2: Sum over the colour of one distinguished vertex.}
We single out \(v_1\) and choose
\begin{equation}
\lambda_1(j)=1
\qquad
\text{for every }j\in C_4.
\label{eq:lambda-one-improved}
\end{equation}
For the three possible partners of \(v_1\), define
\begin{equation}
A_{2j}=\sum_{i\in C_4}W_{12}^{ij},
\quad
A_{3j}=\sum_{i\in C_4}W_{13}^{ij},
\quad
A_{4j}=\sum_{i\in C_4}W_{14}^{ij}.
\label{eq:A-four-vertex-improved}
\end{equation}
These entries form the \(3\times4\) matrix
\begin{equation}
A=
\begin{pmatrix}
A_{20}&A_{21}&A_{22}&A_{23}\\
A_{30}&A_{31}&A_{32}&A_{33}\\
A_{40}&A_{41}&A_{42}&A_{43}
\end{pmatrix}
\label{eq:n4-row-matrix}
\end{equation}
The rows correspond to the three possible partners \(v_2,v_3,v_4\) of \(v_1\), and the columns to the four colours inherited by that partner. For instance, \(A_{2j}\) sums the effective weights between \(v_1\) and \(v_2\) over all colours at \(v_1\), with colour \(j\) fixed at \(v_2\).

For \(r=2,3,4\), define the row linear functional \(R_r:\mathbb C^4\to\mathbb C\) by
\begin{equation}
R_r(\lambda)
:=
\sum_{j\in C_4}A_{rj}\lambda(j).
\label{eq:four-vertex-row-equations-improved}
\end{equation}
Since \(\lambda_1\equiv1\), the three contractions involving \(v_1\) satisfy \(B_{1r}=R_r(\lambda_r)\), so
\begin{equation}
\boxed{
\mathcal F(\omega)
=
R_2(\lambda_2)B_{34}+R_3(\lambda_3)B_{24}+R_4(\lambda_4)B_{23}.
}
\label{eq:Fomega-four-row-form}
\end{equation}
To obtain \(\mathcal F(\omega)=0\), it is enough to choose \(\lambda_2,\lambda_3,\lambda_4\) so that
\begin{equation}
R_2(\lambda_2)=R_3(\lambda_3)=R_4(\lambda_4)=0,
\label{eq:n4-row-zero-condition}
\end{equation}
Each of the three summands contains one of these vanishing row contractions, so all three are zero regardless of \(B_{34},B_{24},B_{23}\). Hence
\begin{equation}
\mathcal F(\omega)=0.
\label{eq:n4-graph-annihilation}
\end{equation}
Physically, the three row equations enforce destructive interference separately in the three mutually exclusive channels in which \(v_1\) is paired with \(v_2\), \(v_3\), or \(v_4\).

\paragraph*{Step 3: See what the same functional does to the GHZ target.}
The target array \(\Delta\) is nonzero only for the four monochromatic colourings
\begin{equation}
(0,0,0,0),\ (1,1,1,1),\ (2,2,2,2),\ (3,3,3,3).
\label{eq:n4-monochromatic-colourings}
\end{equation}
With \(\lambda_1\equiv1\), summing these monochromatic colourings gives
\begin{equation}
\mathcal F(\Delta)
=
Q_4(\lambda_2,\lambda_3,\lambda_4)
:=
\sum_{j=0}^{3}
\lambda_2(j)\lambda_3(j)\lambda_4(j).
\label{eq:four-vertex-delta-function-improved}
\end{equation}
The four-vertex proof therefore reduces to one question:
\begin{quote}
Can we choose coefficient vectors \(\lambda_r\in\mathbb C^4\), for \(r=2,3,4\), such that the row contractions \(\sum_{j=0}^{3}A_{rj}\lambda_r(j)\) vanish for every \(r\), while keeping the diagonal cubic form \(Q_4\) nonzero?
\end{quote}
The following lemma shows that such a choice always exists.

\begin{lemma}[The \(3\times4\) lemma]
For every complex \(3\times4\) matrix \(A\), there exist vectors
\(\lambda_2,\lambda_3,\lambda_4\in\mathbb C^4\) such that
\begin{equation}
\sum_{j=0}^{3}A_{rj}\lambda_r(j)=0
\qquad
(r=2,3,4),
\label{eq:n4-lemma-row-conditions}
\end{equation}
but
\begin{equation}
Q_4(\lambda_2,\lambda_3,\lambda_4)
=
\sum_{j=0}^{3}
\lambda_2(j)\lambda_3(j)\lambda_4(j)\neq0.
\label{eq:n4-lemma-diagonal-condition}
\end{equation}
\end{lemma}

\begin{proof}
For \(j=0,1,2,3\), let \(e_j\in\mathbb C^4\) have an entry of \(1\) at colour \(j\) and \(0\) elsewhere.

\emph{Case 1: \(A\) has a zero column.}
Suppose column \(j_\star\) is identically zero.  Choose all three vectors to be the one-hot vector at that colour:
\begin{equation}
\lambda_2=\lambda_3=\lambda_4=e_{j_\star}.
\label{eq:n4-zero-column-vectors}
\end{equation}
Each row contraction is
\begin{equation}
\sum_{j=0}^{3}A_{rj}\lambda_r(j)=A_{rj_\star}=0,
\label{eq:n4-zero-column-row-cancellation}
\end{equation}
whereas the diagonal form is
\begin{equation}
Q_4(\lambda_2,\lambda_3,\lambda_4)=1.
\label{eq:n4-zero-column-diagonal-value}
\end{equation}
\emph{Case 2: no column of \(A\) is zero.}
Call a colour \(j\) \emph{blocked} if some row of \(A\) is supported only in column \(j\); that is, for some \(r\),
\begin{equation}
A_{rj}\neq0,
\qquad
A_{rk}=0\quad(k\neq j).
\label{eq:n4-blocked-colour-condition}
\end{equation}
Such a row would force \(\lambda_r(j)=0\), so the diagonal product for a blocked colour must be zero. A single row can block at most one colour. Since there are three rows but four colours, at least one colour is not blocked. Fix one and call it \(j_\star\).

We now treat the rows independently.

If \(A_{rj_\star}=0\), choose
\begin{equation}
\lambda_r=e_{j_\star}.
\label{eq:n4-one-hot-row}
\end{equation}
Then the row contraction is zero, and \(\lambda_r\) is supported only at \(j_\star\).

If \(A_{rj_\star}\neq0\), then the row cannot be supported only at \(j_\star\), because \(j_\star\) is not blocked.  Hence there is an auxiliary colour \(q_r\neq j_\star\) with \(A_{rq_r}\neq0\).  Choose
\begin{equation}
\lambda_r
=
A_{rq_r}e_{j_\star}-A_{rj_\star}e_{q_r}.
\label{eq:n4-two-entry-row}
\end{equation}
The two nonzero entries cancel in the row contraction:
\begin{align}
\sum_{j=0}^{3}A_{rj}\lambda_r(j)
=
A_{rj_\star}A_{rq_r}
+A_{rq_r}(-A_{rj_\star})
=0.
\label{eq:n4-two-entry-cancellation}
\end{align}
Moreover,
\begin{equation}
\lambda_r(j_\star)=A_{rq_r}\neq0.
\label{eq:n4-special-colour-nonzero}
\end{equation}
The remaining requirement is to ensure that no diagonal term with \(j\neq j_\star\) is nonzero. If at least one row obeys \(A_{rj_\star}=0\), its one-hot vector has \(\lambda_r(j)=0\) for every \(j\neq j_\star\), so every other diagonal product contains a zero factor. Otherwise all three rows use the two-entry construction. Since \(A_{rj_\star}\neq0\) for every row and \(j_\star\) is unblocked, each row has at least one admissible auxiliary colour \(q_r\neq j_\star\) with \(A_{rq_r}\neq0\). We claim that the auxiliary colours \(q_2,q_3,q_4\) can be chosen so that they are not all equal.

Suppose instead that every admissible triple \((q_2,q_3,q_4)\) has equal components. Then each row has the same single available auxiliary colour, say \(q\). Every row of \(A\) is therefore supported only in the two columns \(\{j_\star,q\}\). Because there are four colours, choose
\begin{equation}
p\in C_4\setminus\{j_\star,q\}.
\label{eq:n4-unused-colour-choice}
\end{equation}
Then \(A_{2p}=A_{3p}=A_{4p}=0\), so column \(p\) would be zero, contradicting the assumption of Case 2. Thus we can choose \(q_2,q_3,q_4\) not all equal.

Now evaluate the diagonal form.  At \(j_\star\), every factor is nonzero: it is either \(1\) in the one-hot case or \(A_{rq_r}\neq0\) in the two-entry case.  Therefore
\begin{equation}
\lambda_2(j_\star)\lambda_3(j_\star)\lambda_4(j_\star)\neq0.
\label{eq:n4-special-colour-product}
\end{equation}
For every colour \(j\neq j_\star\), at least one coefficient \(\lambda_r(j)\) is zero. This is immediate if some \(\lambda_r\) is one-hot at \(j_\star\). If all three functions use auxiliary colours, the fact that \(q_2,q_3,q_4\) are not all equal means that no single \(j\neq j_\star\) equals all three auxiliary colours; hence, for some \(r\), the function \(\lambda_r\) is supported on \(\{j_\star,q_r\}\) with \(q_r\neq j\). Thus
\begin{equation}
\lambda_2(j)\lambda_3(j)\lambda_4(j)=0
\qquad(j\neq j_\star).
\label{eq:n4-other-colour-products}
\end{equation}
Only the \(j_\star\)-term remains in the diagonal sum, so
\begin{equation}
Q_4(\lambda_2,\lambda_3,\lambda_4)
=
\lambda_2(j_\star)\lambda_3(j_\star)\lambda_4(j_\star)
\neq0.
\label{eq:n4-diagonal-form-nonzero}
\end{equation}
The row contractions vanish while \(Q_4\) remains nonzero, proving the lemma.
\end{proof}

\paragraph*{Step 4: Numerical examples of both cases.}
\emph{A zero column.} Consider
\begin{equation}
A=\begin{pmatrix}
0&1&2&0\\
0&3&0&4\\
0&0&5&6
\end{pmatrix}.
\label{eq:n4-zero-column-example-matrix}
\end{equation}
The first column is zero, so choose \(\lambda_2=\lambda_3=\lambda_4=(1,0,0,0)\). All three row contractions are zero, while \(Q_4=1\cdot1\cdot1=1\).

\emph{No zero column.} The auxiliary colours must be chosen carefully. Consider
\begin{equation}
A=
\begin{pmatrix}
1&1&2&0\\
1&1&0&3\\
1&1&4&5
\end{pmatrix},
\qquad
j_\star=0.
\label{eq:n4-auxiliary-example-matrix}
\end{equation}
Suppose every row uses the same auxiliary colour \(q_r=1\). Then
\begin{equation}
\lambda_2=\lambda_3=\lambda_4=(1,-1,0,0).
\label{eq:n4-common-auxiliary-vectors}
\end{equation}
These vectors make all three row contractions vanish, but
\begin{equation}
Q_4(\lambda_2,\lambda_3,\lambda_4)=1\cdot1\cdot1+(-1)(-1)(-1)=0.
\label{eq:n4-common-auxiliary-cancellation}
\end{equation}
The \(j=1\) diagonal term cancels the \(j=0\) term.

Instead choose
\begin{equation}
q_2=2,
\qquad
q_3=3,
\qquad
q_4=1.
\label{eq:n4-distinct-auxiliary-colours}
\end{equation}
The two-entry construction in Eq.~\eqref{eq:n4-two-entry-row} then gives
\begin{align}
\lambda_2&=(2,0,-1,0),
\nonumber\\
\lambda_3&=(3,0,0,-1),
\nonumber\\
\lambda_4&=(1,-1,0,0).
\label{eq:n4-distinct-auxiliary-vectors}
\end{align}
The row contractions still vanish, but now
\begin{equation}
Q_4(\lambda_2,\lambda_3,\lambda_4)=2\cdot3\cdot1=6\neq0.
\label{eq:n4-distinct-auxiliary-value}
\end{equation}
Here the \(j=0\) diagonal product is nonzero, while every other diagonal product contains a zero factor.

\paragraph*{Step 5: The contradiction for \(N=D=4\).}
The \(3\times4\) lemma constructs \(\lambda_2,\lambda_3,\lambda_4\) for the matrix obtained from \emph{any} proposed four-vertex graph. Together with \(\lambda_1\equiv1\), Eq.~\eqref{eq:n4-graph-annihilation} gives \(\mathcal F(\omega)=0\), while Eqs.~\eqref{eq:four-vertex-delta-function-improved} and \eqref{eq:n4-diagonal-form-nonzero} give \(\mathcal F(\Delta)\neq0\). This contradicts the functional equality forced by the assumed array identity in Eq.~\eqref{eq:n4-assumed-table-equality}. Therefore no complex-weighted monochromatic quantum graph exists for \(N=4,D=4\).

\subsection*{Generalisation to arbitrary even \texorpdfstring{$N$}{N}}
The four-vertex proof already contains every conceptual ingredient.  The general case changes only the bookkeeping:
\begin{itemize}
\item the three perfect vertex pairings are replaced by all elements of \(\mathcal{PM}(K_N)\);
\item the three possible partners of \(v_1\) are replaced by \(N-1\) possible partners;
\item the \(3\times4\) matrix becomes an \((N-1)\times N\) matrix; and
\item the inequality \(3<4\) becomes \(N-1<N\).
\end{itemize}
Return to the square case \(D=N\) and the colour set \(C_N\) defined in Eq.~\eqref{eq:general-colour-set}. As in Eq.~\eqref{eq:lambda-one-improved}, choose
\begin{equation}
\lambda_1(j)=1
\qquad
\text{for every }j\in C_N.
\label{eq:general-lambda-one}
\end{equation}
The general case uses the same factorisation idea as before: every perfect vertex pairing contains exactly one pair containing each vertex. For \(P\in\mathcal{PM}(K_N)\), generalise the fixed-matching contribution on the left-hand side of Eq.~\eqref{eq:first-matching-factorization-improved} by defining
\begin{equation}
\mathcal F_P(\omega)
:=
\sum_{\mathbf c\in C_N^N}
\left(
\prod_{t=1}^N \lambda_t(c_{v_t})
\right)
\prod_{\substack{\{v_r,v_s\}\in P\\ r<s}}
W_{rs}^{c_{v_r}c_{v_s}}.
\label{eq:fixed-pairing-contribution}
\end{equation}
Because \(P\) partitions the vertices into disjoint pairs, each colour variable occurs in exactly one pair factor. Separating the colour sums as in Eq.~\eqref{eq:first-matching-factorization-improved} gives
\begin{align}
\mathcal F_P(\omega)
&=\prod_{\substack{\{v_r,v_s\}\in P\\ r<s}}
\left(
\sum_{a,b\in C_N}
\lambda_r(a)\lambda_s(b)W_{rs}^{ab}
\right).
\label{eq:general-factorization-without-B}
\end{align}
Summing all matching contributions generalises the three-term expression in Eq.~\eqref{eq:Fomega-four-explicit}:
\begin{equation}
\mathcal F(\omega)
=
\sum_{P\in\mathcal{PM}(K_N)}\mathcal F_P(\omega).
\label{eq:functional-sum-over-pairings}
\end{equation}

Now use the special role of \(v_1\). Every perfect vertex pairing contains exactly one pair \(\{v_1,v_r\}\), where \(r\in\{2,\ldots,N\}\). Extending the effective-weight sums in Eq.~\eqref{eq:A-four-vertex-improved}, define for \(r=2,\ldots,N\) and \(j\in C_N\)
\begin{equation}
A_{rj}=\sum_{i\in C_N}W_{1r}^{ij}.
\label{eq:definition-A-improved}
\end{equation}
The row index \(r\) records the possible partner \(v_r\) of \(v_1\), and the column index \(j\) records the colour inherited by that partner. Thus \(A\) is the \((N-1)\times N\) counterpart of the matrix in Eq.~\eqref{eq:n4-row-matrix}. For \(r=2,\ldots,N\), extend the row functionals in Eq.~\eqref{eq:four-vertex-row-equations-improved} by defining \(R_r:\mathbb C^{C_N}\to\mathbb C\) as
\begin{equation}
R_r(\lambda)
:=
\sum_{j\in C_N}A_{rj}\lambda(j).
\label{eq:general-row-functional}
\end{equation}

Since \(\lambda_1\equiv1\), the same reduction that led to the four-vertex row form in Eq.~\eqref{eq:Fomega-four-row-form} identifies the factor for \(\{v_1,v_r\}\) with its row contraction:
\begin{equation}
\begin{aligned}
\sum_{i,j\in C_N}\lambda_1(i)\lambda_r(j)W_{1r}^{ij}
&=\sum_{i,j\in C_N}\lambda_r(j)W_{1r}^{ij}\\
&=\sum_{j\in C_N}
\left(\sum_{i\in C_N}W_{1r}^{ij}\right)\lambda_r(j)\\
&=R_r(\lambda_r).
\end{aligned}
\label{eq:distinguished-edge-row-contraction}
\end{equation}
Therefore, if we impose the row cancellations from Eq.~\eqref{eq:n4-row-zero-condition} for all possible partners,
\begin{equation}
R_r(\lambda_r)=0
\qquad \text{for every }r=2,\ldots,N,
\label{eq:row-equations-improved}
\end{equation}
then every matching contribution has a zero factor from its pair containing \(v_1\). Summing over the matchings gives the same vanishing as in Eq.~\eqref{eq:n4-graph-annihilation}:
\begin{equation}
\mathcal F(\omega)=0.
\label{eq:graph-side-zero-general}
\end{equation}

On the target side, setting \(\lambda_1\equiv1\) replaces the cubic form \(Q_4\) in Eq.~\eqref{eq:four-vertex-delta-function-improved} by
\begin{equation}
\mathcal F(\Delta)
=Q_N(\lambda_2,\ldots,\lambda_N)
:=\sum_{j\in C_N}\prod_{r=2}^N\lambda_r(j).
\label{eq:delta-after-lambda-one-improved}
\end{equation}
It remains to satisfy the row equations while keeping \(Q_N\) nonzero. The following lemma extends the two conditions in Eqs.~\eqref{eq:n4-lemma-row-conditions} and \eqref{eq:n4-lemma-diagonal-condition} to \(N-1\) rows and \(N\) colours.
\begin{lemma}[The \((N-1)\times N\) lemma]
Let \(N\geq4\), let \(C_N\) be the colour set in Eq.~\eqref{eq:general-colour-set}, and let \(A\) be an
\((N-1)\times N\) complex matrix with rows indexed by \(r=2,\ldots,N\)
and columns indexed by \(j\in C_N\). With the row functionals \(R_r\) from Eq.~\eqref{eq:general-row-functional}, there exist functions
\(\lambda_r:C_N\to\mathbb{C}\), for \(r=2,\ldots,N\), such that
\begin{equation}
R_r(\lambda_r)=0
\qquad\text{for every }r=2,\ldots,N,
\label{eq:matrix-lemma-row-equations}
\end{equation}
but
\begin{equation}
Q_N(\lambda_2,\ldots,\lambda_N)\neq0.
\label{eq:matrix-lemma-diagonal-nonzero}
\end{equation}
\end{lemma}

The proof uses exactly the construction established for the \(3\times4\) lemma; only the index ranges and the row--column count change.

\begin{proof}
For \(j\in C_N\), let \(e_j:C_N\to\mathbb C\) be the coordinate function with \(e_j(j)=1\) and \(e_j(k)=0\) for \(k\neq j\).

\emph{Zero column.}
If column \(j_\star\) of \(A\) is zero, set \(\lambda_r=e_{j_\star}\) for every \(r=2,\ldots,N\). This is the construction in Eqs.~\eqref{eq:n4-zero-column-vectors}--\eqref{eq:n4-zero-column-diagonal-value}, with three row vectors replaced by \(N-1\): every row satisfies \(R_r(\lambda_r)=A_{rj_\star}=0\), while \(Q_N(\lambda_2,\ldots,\lambda_N)=1\).

\emph{Blocked and unblocked columns.}
Suppose that \(A\) has no zero column. As in Eq.~\eqref{eq:n4-blocked-colour-condition}, call a column blocked if some row is supported only in that column. Each of the \(N-1\) rows blocks at most one column, whereas \(A\) has \(N\) columns. Hence an unblocked column \(j_\star\) exists. This replaces the inequality \(3<4\) in the four-vertex proof by \(N-1<N\).

\emph{One-hot and two-entry functions.}
For each \(r=2,\ldots,N\), use the same row-wise definitions as in Eqs.~\eqref{eq:n4-one-hot-row} and \eqref{eq:n4-two-entry-row}. If \(A_{rj_\star}=0\), set \(\lambda_r=e_{j_\star}\). If \(A_{rj_\star}\neq0\), the fact that \(j_\star\) is unblocked provides a colour \(q_r\neq j_\star\) with \(A_{rq_r}\neq0\); set \(\lambda_r=A_{rq_r}e_{j_\star}-A_{rj_\star}e_{q_r}\). The cancellation in Eq.~\eqref{eq:n4-two-entry-cancellation} is unchanged, so \(R_r(\lambda_r)=0\) in both cases. Moreover, \(\lambda_r(j_\star)\) is nonzero for every row: it equals \(1\) in the one-hot case and \(A_{rq_r}\) in the two-entry case.

\emph{Auxiliary colours.}
If one of the functions is one-hot at \(j_\star\), every diagonal product with \(j\neq j_\star\) already contains a zero factor. Otherwise every row uses the two-entry construction. For each row let \(S_r:=\{q\in C_N\setminus\{j_\star\}:A_{rq}\neq0\}\). Since \(A_{rj_\star}\neq0\) for every row and \(j_\star\) is unblocked, each \(S_r\) is non-empty. The choices \(q_r\in S_r\) can be made not all equal. Indeed, if every admissible tuple had equal components, all the sets \(S_r\) would have to be the same singleton, say \(\{q\}\). Since \(N\geq4\), one could choose \(p\in C_N\setminus\{j_\star,q\}\); then \(A_{rp}=0\) for every \(r\), contradicting the assumption that \(A\) has no zero column. This is the general version of the argument surrounding Eq.~\eqref{eq:n4-unused-colour-choice}.

Choose the \(q_r\) not all equal. For each fixed \(j\neq j_\star\), some row has \(q_r\neq j\), and the corresponding function is supported on \(\{j_\star,q_r\}\). Hence \(\lambda_r(j)=0\) for that row. Thus every off-\(j_\star\) diagonal product is zero, whereas \(\prod_{r=2}^N\lambda_r(j_\star)\neq0\). Only the \(j_\star\)-term remains in \(Q_N\), so \(Q_N(\lambda_2,\ldots,\lambda_N)\neq0\). This proves the lemma.
\end{proof}

Applying the lemma to the graph matrix \(A\) gives \(\mathcal F(\omega)=0\) and \(\mathcal F(\Delta)\neq0\), contradicting \(\omega=\Delta\) and completing the proof for \(D=N\).

\begin{remark}
Because the construction of \(\lambda_r\) uses only addition, subtraction, and multiplication (avoiding division in Eq.~\eqref{eq:n4-two-entry-row}), and the final step \(\prod_{r=2}^N \lambda_r(j_\star)\neq0\) needs only the absence of zero divisors, the proof of \(\omega\neq\Delta\) also works when the weights take values in any commutative integral domain (including finite fields such as \(\mathbb F_2\)).
\end{remark}

\subsection*{Extension from \texorpdfstring{$D=N$}{D = N} to \texorpdfstring{$D>N$}{D > N}}

Deleting every edge that carries a chosen colour at either endpoint preserves all graph amplitudes on the remaining colours, reducing a \(D\)-dimensional GHZ state to dimension \(D-1\). Repeating this for any \(D>N\) would produce an \(N\)-dimensional GHZ state on the same \(N\) vertices, contradicting the case \(D=N\).

\subsection*{Limitation for \texorpdfstring{$D<N$}{D < N}}

The row--column argument does not extend directly to \(D\leq N-1\), and the first boundary case \(N=4,D=3\) shows that the threshold is meaningful. For \(C_3=\{0,1,2\}\), take the six nonzero effective weights
\begin{equation}
\begin{aligned}
W_{12}^{00}=W_{34}^{00}&=1,\\
W_{13}^{11}=W_{24}^{11}&=1,\\
W_{14}^{22}=W_{23}^{22}&=1,
\end{aligned}
\label{eq:three-colour-effective-weights}
\end{equation}
with all other effective weights equal to zero. The three possible vertex pairings produce, respectively, the colourings \((0,0,0,0)\), \((1,1,1,1)\), and \((2,2,2,2)\). These are therefore the only nonzero amplitudes, each with value \(1\), so \(\omega=\Delta\).

After setting \(\lambda_1\equiv1\), the matrix formed from the edges incident to \(v_1\) is
\begin{equation}
A=
\begin{pmatrix}
1&0&0\\
0&1&0\\
0&0&1
\end{pmatrix}
=I_3.
\label{eq:three-colour-incidence-matrix}
\end{equation}
Thus the row-cancellation conditions and their effect on the target are
\begin{equation}
\begin{gathered}
\lambda_2(0)=\lambda_3(1)=\lambda_4(2)=0,\\
\Longrightarrow\quad
\mathcal F(\Delta)
=\sum_{j=0}^{2}\prod_{r=2}^{4}\lambda_r(j)
=0.
\end{gathered}
\label{eq:three-colour-separator-obstruction}
\end{equation}
Each target summand vanishes because a different row contributes a zero factor. Hence the row-wise separator annihilates the target together with the graph amplitude array. The matrix is \(3\times3\), so there is no additional column from which to choose an unblocked colour; this is precisely where the strict row--column inequality fails.

\section*{Physical interpretation}
We now aim to interpret the proof physically. For simplicity, we focus on the case $N=D=4$, which naturally generalises.

In the first step, we observe that $\omega=\Delta$ leads to $\mathcal F(\omega)=\mathcal F(\Delta)$. Physically, this means that if a construction from the graph (i.e., $\omega$) is equivalent to the GHZ state (i.e., $\Delta$), then any set of measurements performed on both must give the same result, i.e., $\mathcal F(\omega)=\mathcal F(\Delta)$.

The proof constructs a set of measurements for which the GHZ state and any state generated from the weighted graph give different results.

The first measurement projects photon~1 onto the equal superposition $\ket{+_4}=(\ket{0}+\ket{1}+\ket{2}+\ket{3})/2$, corresponding to $\lambda_1(j)=1$ in Eq.~\eqref{eq:lambda-one-improved}, apart from normalisation.

This gives the matrix $A$ a physical meaning: within the contribution in which photon~1 is paired with path~$r$, the projection leaves its partner with an amplitude $A_{rj}/2$ in internal mode $j$ (such as colour).
 
Thus, each row of $A$ describes a partner-photon spatial path, while each column labels an internal mode [Eq.~\eqref{eq:A-four-vertex-improved}]. As the measurement of photon 1 is in a coherent superposition, the three pairing contributions remain coherent. The measurement does not reveal which photon was the partner.

Now the matrix $A$ is analysed in two cases -- both of which show that the results for the GHZ state must differ from those for states generated from the graphs.

In Case~1, a zero column $j_\star$ means that, after this projection, photon~1's partner has zero amplitude to occupy internal mode $j_\star$, irrespective of its spatial path. The coincidence $(+_4,j_\star,j_\star,j_\star)$ selected in Eq.~\eqref{eq:n4-zero-column-vectors} is therefore forbidden: detecting all three remaining photons in that mode would necessarily require the partner to occupy it. The normalised GHZ target, however, yields this same outcome with probability $1/16$, distinguishing it from the proposed pair-generated state.

In Case~2, no column is zero, so there is no internal mode that is absent for every possible partner path. Instead, we choose local superposition measurements that suppress each partner contribution by destructive interference. With four modes but only three possible partners, there must be a mode $j_\star$ that is not the sole nonzero mode of any partner contribution. This allows us to retain $j_\star$ in every local measurement while cancelling the corresponding partner amplitude.

For each possible partner \(r\) of photon 1, we choose a local measurement that makes the contribution from that pairing vanish. If \(A_{rj_\star}=0\), projecting onto mode \(j_\star\) already gives zero amplitude. Otherwise, we project onto a superposition of \(j_\star\) and a second mode \(q_r\), choosing their weights so that the two contributions cancel by destructive interference. Thus, whichever photon is paired with photon 1, that pair gives zero amplitude for the selected outcome, regardless of the other pair. The joint coincidence is therefore forbidden.

However, all three local measurements retain a nonzero overlap with mode \(j_\star\), and we choose them so that no other mode is shared by all three. Since each GHZ term has all photons in the same mode, only the term \(\lvert j_\star\rangle^{\otimes4}\) contributes to this coincidence, with nonzero amplitude. The same outcome is therefore impossible for the proposed pair-generated state but possible for the GHZ target.

As a consequence, we have seen that there cannot be any graph with weights $\omega$ that correspond to a $N=D=4$ GHZ state. The generalisation to higher dimensions follows the same principles.

\section*{Conclusion and Outlook}
We have proven that high-dimensional GHZ states with $N$ particles and $D$ dimensions cannot be generated with minimal resources (linear optics, pair-creation sources, and no ancillas or heralding). This answers, for the first time, a general case of a question that we raised in 2017 \cite{krenn2017entanglement, krenn2017quantum}. The answer has been discovered by an AI system called \textit{AlphaProof Nexus} \cite{tsoukalas2026advancing}. The underlying mathematics is surprisingly simple. Despite the enormously complex space of potential experiments, rather simple mathematical objects (tensors), combined with a number of clever tricks, were able to solve the question in the end. The simplicity of the underlying mathematics also allows for a physical interpretation of the proof and all of the objects in the proof.

Since the formalized proof of the theorem presented in this manuscript was discovered and published online, several other special cases and several important extensions have been discovered -- most importantly, five months after the discovery of our proof, Rishi Rajasekaran, using \textit{Codex}, discovered a proof of the entire conjecture \cite{rajasekaran2026krenngu}. The proof has been formally verified via Lean. The core idea is based on a related technique that we present here (finding an $f(\omega)\neq f(\Delta)$ while $\omega = \Delta$). However -- to the best of our knowledge, no human has digested and understood the proof yet, nor translated the insights into quantum optics. This will be a very important, immediate next step. 

Given that the conjecture has been solved for very large classes of cases (and fully solved afterwards), it may be useful to consider the physical consequences and extensions. We now know that high-dimensional GHZ states for larger quantum systems cannot be generated with the minimal resources possible (e.g., no heralding and no ancillary photons) using linear optics and photon pair creation. This immediately raises two questions: (1) Which quantum states can be generated in this way? We now know that there are definite limits, but a full characterization would be very helpful. (2) Can higher-order photon creation processes, such as photon-triple creation processes, be used to generate arbitrarily high-dimensional multiphotonic GHZ states? It has been shown that high-dimensional multiphoton GHZs can be generated in this way \cite{gu2020quantum}, but it is unknown what the limits are. Both of these questions can be formulated in precise mathematical terms and formalized using, again, the language of graphs (hypergraphs) and perfect matchings, and will be useful extensions to the \textit{Formal Conjectures} project \cite{FormalConjectures}.

This whole research program points towards an exciting possibility, namely learning from AI agents for mathematics in physics. Specifically, in this case, a key step was to translate the physics question into a mathematical question \cite{krenn2017quantum}, then formalise the mathematical question in Lean \cite{FormalConjectures, FormalConjecturesPaper}, subsequently solve it with AI \cite{tsoukalas2026advancing}, and finally provide -- as we do here -- a full description and physical interpretation of the proof. We expect that now, in the era of AI-assisted mathematical proofs, we will see many more results in physics being achieved once the corresponding questions are fully described mathematically and formalised \cite{krippendorf2026future}.

\section*{Acknowledgements}
MK acknowledges support from the European Research Council (ERC) under the European Union's Horizon Europe research and innovation programme (ERC-2024-STG, Grant No.~101165179, ArtDisQ) and from the German Research Foundation (DFG; EXC 2064/1, Project 390727645). X.G. acknowledges support from the Alexander von Humboldt Foundation and the Deutsche Forschungsgemeinschaft (DFG, German Research Foundation), Project ID 398816777 (SFB 1375).

\bibliography{refs}

\end{document}